\documentclass[12pt]{article}
\pdfoutput=1
\usepackage[a4paper]{geometry}
\usepackage{jheppub, amsmath,amssymb,amsfonts,amsxtra, mathrsfs, makeidx,graphics,graphicx,amsthm,epsfig, ytableau,bm,longtable,float, color,tikz,mathtools,xfrac,footnote,rotating, lscape, makecell}
\usetikzlibrary{positioning}
\usetikzlibrary{chains}
\usetikzlibrary{arrows,fit,decorations.pathreplacing}
\tikzstyle{every picture}+=[remember picture]
\tikzstyle{na} = [baseline]

\usetikzlibrary{arrows, decorations.markings, calc, fadings, decorations.pathreplacing, patterns, decorations.pathmorphing, positioning}

\def\node#1#2{\overset{#1}{\underset{#2}{{\color{gray} \bullet}}}}
\def\Node#1#2{\overset{#1}{\underset{#2}{{ \bullet}}}}
\def\NNode#1#2{\overset{#1}{\underset{#2}{{\color{blue} \bullet}}}}

\def\ver#1#2{\overset{{\llap{$\scriptstyle#1$}\displaystyle{\color{gray} \blacksquare}{\rlap{$\scriptstyle#2$}}}}{\scriptstyle\vert}}
\def\vver#1#2{\overset{{\llap{$\scriptstyle#1$}\displaystyle{\color{red} \blacksquare}{\rlap{$\scriptstyle#2$}}}}{\scriptstyle\vert}}
\def\Ver#1#2{\overset{{\llap{$\scriptstyle#1$}\displaystyle\blacksquare{\rlap{$\scriptstyle#2$}}}}{\scriptstyle\vert}}

\usepackage{bbm}
\usepackage[vcentermath]{youngtab}
\usepackage{subfigure}

\newcommand{\eg}{\textit{e.g.}}

\newcommand{\ie}{\textit{i.e.}}

\numberwithin{equation}{section}

\newcommand{\nn}{\nonumber}

\newcommand{\be}{\begin{equation}} \newcommand{\ee}{\end{equation}}
\newcommand{\bea}{\begin{equation} \begin{aligned}} \newcommand{\eea}{\end{aligned} \end{equation}}

\def\hat{\widehat}

\def\rt2{\sqrt{2}}

\def\tr{\mathop{\rm tr}}

\def\CO{{\cal O}}

\def\1{{\ds 1}}

\newcommand{\bR}{\mathbb{R}}
\newcommand{\bZ}{\mathbb{Z}}

\newcommand{\fn}{\mathfrak{n}}

\def\repa{\raise4pt\hbox{$\square$}\mkern-14mu\raise-4pt\hbox{$\square$}}
\def\repab{\overline{\raise4pt\hbox{$\square$}\mkern-14mu\raise-4pt\hbox{$\square$}\mkern-1mu}}

\def\smileface{\ensuremath{\hbox{\large$\bigcirc$}\mkern-15mu\raise-1pt\hbox{\scriptsize$\smallsmile$}%
\mkern-10mu\raise4pt\hbox{..}\mkern4mu}}
\def\frownface{\ensuremath{\hbox{\large$\bigcirc$}\mkern-15mu\raise-1pt\hbox{\scriptsize$\smallfrown$}%
\mkern-10mu\raise4pt\hbox{..}\mkern4mu}}

\newcommand{\ba}{\begin{array}}
\newcommand{\ea}{\end{array}}
\newcommand{\bi}{\begin{itemize}}
\newcommand{\ei}{\end{itemize}}
\def\vec#1{\bm{#1}}
\def\bea#1\eea{\allowdisplaybreaks \begin{align}#1\end{align}}
 \newcommand{\ben}{\begin{enumerate}}
\newcommand{\een}{\end{enumerate}}
\newcommand{\bean}{\begin{eqnarray*}}
\newcommand{\eean}{\end{eqnarray*}}
\newcommand{\eref}[1]{(\ref{#1})}

\newcommand{\BC}{\mathbb{C}}

\newcommand{\BZ}{\mathbb{Z}}

\newcommand{\comment}[1]{}

\title{T-branes, Anomalies and Moduli Spaces in 6D SCFTs}
\author[a,b]{Noppadol Mekareeya,}
\author[c]{Tom Rudelius,}
\author[a,b]{and Alessandro Tomasiello}

\affiliation[a]{Dipartimento di Fisica, Universit\`a di Milano-Bicocca, \\ Piazza della Scienza 3, I-20126 Milano, Italy}
\affiliation[b]{INFN, sezione di Milano-Bicocca, I-20126 Milano, Italy}
\affiliation[c]{Jefferson Physical Laboratory, Harvard University, \\ Cambridge, MA 02138, USA}

\abstract{The worldvolume theory of M5-branes on an ADE singularity $\bR^5/\Gamma_G$ can be Higgsed in various ways, corresponding to the possible nilpotent orbits of $G$. In the F-theory dual picture, this corresponds to activating T-brane data along two stacks of 7-branes and yields a tensor branch realization for a large class of 6D SCFTs. In this paper, we show that the moduli spaces and anomalies of these T-brane theories are related in a simple, universal way to data of the nilpotent orbits. This often works in surprising ways and gives a nontrivial confirmation of the conjectured properties of T-branes in F-theory. We use this result to formally engineer a class of theories where the IIA picture na\"ively breaks down. We also give a proof of the $a$-theorem for all RG flows within this class of T-brane theories.}

\emailAdd{n.mekareeya@gmail.com}
\emailAdd{rudelius@physics.harvard.edu}
\emailAdd{alessandro.tomasiello@unimib.it}

\begin{document}
\maketitle

\section{Introduction}

M5-branes are one of the most mysterious elements of M-theory. They are described by a six-dimensional superconformal field theory (6D SCFT) \cite{Witten:1995zh,Strominger:1995ac,Witten:1995em} with ${\cal N}=(2,0)$ supersymmetry, chiral tensor gauge fields, and a number of degrees of freedom growing as $N^3$ rather than $N^2$.  A related class of ${\cal N}=(1,0)$ 6D SCFTs is obtained by taking the M5-branes to probe a $\bR \times \bR^4/\Gamma_G$ singularity, where $\Gamma_G$ is an ADE discrete subgroup of $SU(2)$ and $G$ is its McKay-dual Lie group. These orbifold theories are characterized by a number of degrees of freedom that grows like $|\Gamma_G|^2 N^3$ and a $G\times G$ flavor symmetry.  Anomaly cancellation and positivity of the Dirac pairing on the string charge lattice tightly constrain these theories and allow more generally for a systematic classification of 6D SCFTs \cite{Heckman:2013pva, Heckman:2015bfa, Bhardwaj:2015xxa}.

It is sometimes helpful to consider this configuration from dual points of view. For the $G=SU(k)$ case, one can study a IIA reduction, which consists of a NS5--D6 brane intersection. The resulting theory is a chain of $SU(k)$ gauge groups.  For $G=SO(2k)$, one adds O6-planes: the NS5-branes subsequently fractionate, yielding alternating $SO(2k)$ and $Sp(k-4)$ gauge groups connected by bifundamental hypermultiplets \cite{Hanany:1997gh}. On the other hand, for $G=E_k$, a weakly-coupled IIA realization does not exist. Nonetheless, these theories can be constructed using F-theory \cite{DelZotto:2014hpa}, resulting in a linear chain of $G$ gauge groups connected by generalizations of bifundamentals known as ``conformal matter" \cite{DelZotto:2014hpa}.  This conformal matter can be interpreted as M5-brane fractionation, which generalizes the NS5 fractionation mentioned above and was confirmed using string dualities in \cite{ohmori-shimizu-tachikawa-yonekura-T2,tachikawa-frozen}.

The story becomes much richer once one starts exploring the Higgs moduli spaces for these orbifold M5-brane theories. Giving a vev along certain directions of the moduli spaces breaks the flavor symmetry $G\times G$ and triggers an RG flow to a different CFT; different vevs give rise to different endpoints for this flow. The string theory realization for the theories leads to an expectation that this process should be related to the theory of nilpotent orbits. For example, in the IIA realization, one can adapt \cite{Gaiotto:2014lca} a logic used for D3-NS5-brane systems in \cite{Gaiotto:2008sa}. The BPS equations for the scalars transverse to the D6-branes are the Nahm equations, whose boundary conditions are related to partitions and hence to the nilpotent orbits of $G=SU(k)$. (We review this in section \ref{sub:suk}.) In the F-theory picture, a similar conclusion is suggested \cite{DelZotto:2014hpa} by the Hitchin equations living on the 7-branes. This logic can be applied independently to the two factors $G$ of the flavor symmetry. Thus we end up with a set of CFTs labeled by two nilpotent orbits ${\cal O}_{\rm L}$, ${\cal O}_{\rm R}$ of $G$.

Some nilpotent orbits are properly contained in others. This induces a partial ordering that is isomorphic to the ordering under RG flows \cite{Heckman:2016ssk}. Mathematically, for the classical groups, this is also known as the Kraft--Procesi transition \cite{kraftprocesi} (see also \cite{Cabrera:2016vvv} for the discussion from the brane perspective).  In the $G=SU(k)$ case, the partially Higgsed theories described above can be constructed in IIA by adding D8-branes \cite{Gaiotto:2014lca,Hanany:1997gh} (see also \cite{afrt,10letter,cremonesi-t} for the AdS$_7$ duals of these theories and also \cite{Cremonesi:2014kwa, Cabrera:2016vvv} for the brane realization of the Kraft--Procesi transition in the context of 3d gauge theories). In the $G=SO(2k)$ case, one may attempt a similar IIA construction with O6-planes and D8-branes, but this breaks down near the bottom of the RG hierarchy, where spinor representations appear \cite{Heckman:2016ssk}.  In this case, as well as the $G=E_k$ case, one must turn to F-theory to construct the Higgsed theories.

The relation of these theories to the nilpotent orbits suggests that there should also be a relation between the dimensions $d_{{\cal O}_{\rm L}}$, $d_{{\cal O}_{\rm R}}$ of the two nilpotent orbits $\CO_{\rm L}$, $\CO_{\rm R}$, and the corresponding Higgs moduli space dimension $d_H(\CO_{\rm L}, \CO_{\rm R})$.  In this paper we argue that this is indeed the case: given a theory labeled by nilpotent orbits $({\cal O}_{\rm L},{\cal O}_{\rm R})$ of group $G$, we have
\bea \label{eq:dim}
d_{H}(\vec{0})-d_H(\CO_{\rm L},\CO_{\rm R}) = d_{{\cal O}_{\rm L}} + d_{{\cal O}_{\rm R}}~,
\eea
where $d_H(\vec{0})$ is the Higgs moduli space of the worldvolume of multiple M5-branes on $\BC^2/\Gamma$, with $\Gamma$ the discrete group that is in the McKay correspondence to $G$. As we will see in several examples (see section \ref{sec:coeffdelta} below), the correspondence (\ref{eq:dim}) works in often quite nontrivial ways, connecting orbit dimensions to the dimensions of various Lie groups and their representations. This is additional strong evidence in favor of a T-brane \cite{cecotti-cordova-heckman-vafa} interpretation of these theories in terms of Hitchin poles, after the correspondence of Hasse diagrams already found in \cite{Heckman:2016ssk}.
 
We will also show that the nilpotent orbit data is related to the anomaly polynomial of the corresponding 6D SCFT, 
\begin{equation}
I = \alpha c_2(R)^2 + \beta c_2(R) p_1(T) + \gamma p_1(T)^2 + \delta p_2(T)
\end{equation}
where $c_2(R)$ represents the second Chern class of the background $SU(2)$ R-symmetry, $p_1(T)$ and $p_2(T)$ represent the Pontryagin classes of the tangent bundle of a formal eight-manifold, and $\alpha$, $\beta$, $\gamma$, and $\delta$ are numerical coefficients. We express the coefficients $\gamma$ and $\delta$ for T-brane theories simply in terms of the dimension of the corresponding nilpotent orbit.  For the theories of type $G= SU(k)$ and $G = SO(2k)$, we write $\alpha$ and $\beta$ in terms of nilpotent orbit data.  For theories of type $G= E_k$, for which the total number of T-brane theories is finite, we derive formulae for $\alpha$ and $\beta$ for each such theory.  Using these formulae, we prove the $a$-theorem among this class of RG flows.

In the case $G=SO(2k)$, we note that the anomaly polynomial for theories with spinor representations, which do not admit a construction in perturbative Type IIA string theory, can nonetheless be obtained via analytic continuation from the anomaly polynomials of ``formal'' IIA diagrams involving negative numbers of branes. A systematic analysis reveals that negative branes can only appear in a small, finite set of formal IIA diagrams. By comparing the moduli space dimensions and anomaly polynomials, we construct a list of how each formal IIA diagram is cured of its negative branes in the correct F-theory picture and thus construct explicitly the T-brane theories of type $G = SO(2k)$ for all $k$, generalizing the work of \cite{Heckman:2016ssk}.

The paper is organized as follows: in section \ref{sec:6D}, we review 6D SCFTs and the T-brane theories in question.  In section \ref{sec:anomaly}, we present formulae that relate the anomaly coefficients $\beta$, $\gamma$, and $\delta$ to nilpotent orbit data universally for all T-brane theories.  In \ref{sec:SO(2n)}, we show how formal IIA diagrams with negative numbers of branes can be used to classify the T-brane theories of type $G=  SO(2k)$ and express each of the anomaly coefficients in terms of these formal IIA diagrams.  In section \ref{sec:athm}, we use these results to prove the $a$-theorem for Higgs branch flows between T-brane theories.  In \ref{sec:conclusions}, we conclude and present directions for future research.  In Appendix \ref{formulae}, we collect the formulae we have derived for all anomaly coefficients.

\section{Six-dimensional SCFTs and T-branes}\label{sec:6D}

In this section we review some aspects of the class of six-dimensional theories of interest. We begin with a general overview in section \ref{sub:gen}. In section \ref{sub:suk} we focus on the theories for $G=SU(k)$, for which the relation of the moduli space dimension to the nilpotent orbit dimension, (\ref{eq:dim}), can be computed directly. In section \ref{sub:anom} we briefly introduce the anomaly polynomial, which we will use in section \ref{sec:anomaly} to show (\ref{eq:dim}) in general. 

\subsection{Overview} 
\label{sub:gen}

Six-dimensional SCFTs are generated by compactifying F-theory on a singular elliptically-fibered Calabi-Yau threefold $X\rightarrow B$ with base $B$ \cite{Heckman:2013pva, Heckman:2015bfa}.  To reach the SCFT point, we simultaneously contract all of the 2-cycles of $B$ to zero size.  Upon blowing up these curves to finite size, thereby moving onto the ``tensor branch" of the 6D theory, we get a smooth base with a collection of 2-cycles intersecting according to some intersection matrix $\Omega$.  The elliptic fiber may degenerate over some of these 2-cycles, leading to gauge symmetries in the corresponding 6D theory.  The allowed degeneration types were classified by Kodaira \cite{Kodaira} and the dictionary between degeneration types and gauge algebras can be found in e.g. \cite{Grassi:2011hq}.  A 6D SCFT is thus characterized by a configuration of 2-cycles with intersection matrix $\Omega$ and a Kodaira type for each 2-cycle.

Conveniently, the intersection matrix of 6D SCFT is tightly constrained by the requirement that all 2-cycles must be simultaneously contractible, and the allowed gauge algebras and matter are tightly constrained by anomaly cancellation.  In particular, no ``loops" are allowed in the configuration of intersecting curves---all 6D SCFTs take the form of a tree-like quiver---and all irreducible curves must have a negative self-intersection number.  We often depict such a quiver by displaying the negative of the self-intersection numbers of each of the curves along with the associated gauge algebras.  For instance, the quiver
\begin{equation}
\overset{\mathfrak{su}(3)}{3} \,\,  1  \,\, \overset{\mathfrak{e}_6}{6}
\end{equation}
represents a configuration in which a curve of self-intersection $-1$ intersects curves of self-intersection $-3$ and $-6$ at one point each.  The gauge algebra $\mathfrak{e}_6$ is associated with the $-6$ curve, and the gauge algebra $\mathfrak{su}(3)$ is associated with the $-3$ curve.

In this paper, we consider a special class of 6D SCFTs that can be realized as deformations of the worldvolume theory of M5-branes probing an orbifold $\mathbb{C}^2/\Gamma_G$.   In F-theory, we write the partial tensor branch for the undeformed theory of type $G$ as
\begin{equation}
\lbrack G] \,\, \underset{\mathfrak{n}}{\underbrace{\overset{\mathfrak{g}}{2} \,\,... \,\,\overset{\mathfrak{g}}{2}}}\,\, [ G ].
\label{eq:typeGquiver}
\end{equation}
Here, the brackets on the left and right indicate the global symmetry group, $G \times G$.  In the case of $G=SO(2k)$ or $E_k$, the F-theory geometry is singular at the points of intersection of these curves and must be blown up via the introduction of curves of self-intersection $-1$, corresponding to the introduction of ``conformal matter."  Thus, for $G=SO(2k)$, the full tensor branch is
\begin{equation}
\lbrack SO(2k)] \,\, \overset{\mathfrak{usp}(2k-8)}{1} \,\,\overset{\mathfrak{so}(2k)}{4} \,\, ...\,\,\overset{\mathfrak{so}(2k)}{4} \,\, \overset{\mathfrak{usp}(2k-8)}{1} \,\, [ SO(2k) ].
\label{eq:SOfull}
\end{equation}
Whereas for $G=E_6$, $E_7$, and $E_8$, it is respectively
\begin{equation}
\lbrack E_6] \,\, {1} \,\,\overset{\mathfrak{su}(3)}{3} \,\,  1  \,\, \overset{\mathfrak{e}_6}{6} \,\, ...\,\,1 \,\,\overset{\mathfrak{su}(3)}{3} \,\, {1}  \,\, [ E_6 ],
\end{equation}
\begin{equation}
\lbrack E_7] \,\, {1} \,\,\overset{\mathfrak{su}(2)}{2} \,\,\overset{\mathfrak{so}(7)}{3} \,\,\overset{\mathfrak{su}(2)}{2} \,\,  1  \,\, \overset{\mathfrak{e}_7}{8} \,\, ... \,\,\overset{\mathfrak{su}(2)}{2} \,\, {1} \,\, [ E_7 ],
\end{equation}
\begin{equation}
\lbrack E_8] \,\, {1}\,\, 2 \,\,\overset{\mathfrak{su}(2)}{2} \,\,\overset{\mathfrak{g}_2}{3} \,\,  1  \,\, \overset{\mathfrak{f}_4}{5} \,\, 1 \,\, \overset{\mathfrak{g}_2}{3}\,\,\overset{\mathfrak{su}(2)}{2} \,\, 2 \,\, 1 \,\,  \overset{\mathfrak{e}_8}{12} \,\, ...\,\,\overset{\mathfrak{su}(2)}{2} \,\, 2 \,\, {1}  \,\, [ E_8 ].
\label{eq:E8full}
\end{equation}
Beginning with these M5-brane theories, we arrive at our ``T-brane theories" of interest by deforming the quivers at the far left and the far right.  The allowed deformations on each side of the quiver of type $G$ have been constructed explicitly and are in one-to-one correspondence with nilpotent orbits of $\Gamma_G$ \cite{Heckman:2016ssk}.  For $G=SU(k)$, these nilpotent orbits are in turn in one-to-one correspondence with partitions of $k$.  For $G=SO(2k)$, they are in one-to-one correspondence with partitions of $2k$ subject to the constraint that any even number appears an even number of times in the partition.

As a first example, we consider the case of $G= SU(3)$, $\Gamma_G = \mathbb{Z}_3$.  The nilpotent orbits of $\mathbb{Z}_3$ are labeled simply by partitions of 3.  The trivial partition $3=1+1+1$ corresponds to the unHiggsed theory
\begin{equation}
	\lbrack SU(3)] \,\, \overset{\mathfrak{su}(3)}{2} \,\,\overset{\mathfrak{su}(3)}{2} \,\,\overset{\mathfrak{su}(3)}{2} \,\,... \,\,\overset{\mathfrak{su}(3)}{2}\,\, [ SU(3) ].
\end{equation}
This theory may be successively Higgsed to the following theories:
\begin{equation}\label{eq:su3flow}
\begin{split}
	\rightarrow  \overset{\mathfrak{su}(2)}{2} \,\,\overset{\mathfrak{su}(3)}{2} \,\,\overset{\mathfrak{su}(3)}{2} \,\,... \,\,\overset{\mathfrak{su}(3)}{2}\,\, [ SU(3) ],\\
	\rightarrow  \overset{\mathfrak{su}(1)}{2} \,\,\overset{\mathfrak{su}(2)}{2} \,\,\overset{\mathfrak{su}(3)}{2} \,\,... \,\,\overset{\mathfrak{su}(3)}{2}\,\, [ SU(3) ].
\end{split}	
\end{equation}
These correspond to the partitions $3=2+1$ and $3=3$, respectively.

Moreover, the RG hierarchy between these Higgsed theories precisely matches the hierarchy between the corresponding nilpotent orbits, which in the case of $G=SU(k)$ or $SO(2k)$ is given by the partial ordering of partitions \cite{Heckman:2016ssk}.  In the present example, the ordering of partitions is given by $1+1+1 \succ 2+1 \succ 3$ and thus matches the RG hierarchy.

More generally, these T-brane theories are characterized by the property that upon blowdown of all $-1$ curves, they consist of a linear chain of $-2$ curves as in (\ref{eq:typeGquiver}).  In fact, they comprise the complete set of theories with this property within the class of known 6D SCFTs, aside from ``outlying" theories with small numbers of tensor multiplets such as\footnote{An ``outlying" theory is more precisely defined as one that doesn't belong to an infinite family of 6D SCFTs.  These infinite families were classified in Appendix B of \cite{Heckman:2015bfa}.}
$$
{1} \,\,\overset{\mathfrak{su}(3)}{3}.
$$

There is more than one way to parametrize the length of our 6D SCFT quivers, and the most convenient way varies depending on the application.  One measure of the length of a quiver is the number of $-2$ curves upon blowing down all $-1$ curves in the F-theory base.  As in (\ref{eq:typeGquiver}), we henceforth denote this quantity as $\mathfrak{n}$.  For the undeformed theories of $Q$ M5-branes probing $\Gamma_G$, $\mathfrak{n}$ is given simply by $Q-1$.  Under flows parametrized by nilpotent deformations, $\mathfrak{n}$ remains constant.  In F-theory terms, this corresponds to the fact that a nilpotent deformation affects the residue of a Higgs field for a Hitchin system on a noncompact ``flavor curve" \cite{Heckman:2016ssk}, but it does not change the number of compact curves in the base \eref{eq:typeGquiver}.

A second measure of the length of a quiver is the total number of tensor multiplets $n_T$, or equivalently the number of curves in the base of the fully-resolved F-theory geometry, as in (\ref{eq:SOfull})-(\ref{eq:E8full}).  For the type $G=SU(k)$ theories, $\mathfrak{n} = n_T$, but for the $G = SO(2k)$ or $G=E_k$ theories this relation no longer holds.

Finally, in the case of $G=SU(k)$, we will often use $N-1 \equiv n_T = \mathfrak{n}$ to denote the length of the quiver.  For $G=SO(2k)$, we will use $N-1 \equiv2 \mathfrak{n}+1$, which is the same as the number of tensor multiplets $n_T$ for the undeformed theory of M5-branes probing a singularity of type $\Gamma_{D_k}$.

\subsection{The $G=SU(k)$ theories}
\label{sub:suk}

We will now review some aspects of the theories with $G=SU(k)$, focusing on their moduli spaces. For more details on the theories themselves and on their holographic duals, see for example \cite{Cremonesi:2015bld}.

We have seen already examples of theories with $G=SU(3)$ in (\ref{eq:su3flow}). In general, the theories with $G=SU(k)$ are all finite chains of gauge groups $SU(r_i)$, $i=1,\ldots,N-1$, such that the maximum of the $r_i$ is $k$. The gauge groups are connected by bifundamental hypermultiplets and are paired with tensor multiplets. The $i$-th gauge group $SU(r_i)$ is also coupled to $f_i$ fundamental hypermultiplets; anomaly cancellation requires
\begin{equation}\label{eq:fi}
	f_i = 2r_i - r_{i+1}-r_{i-1}\ .
\end{equation}
(Since this should be a positive number, the function $i \mapsto r_i$ is convex.) Although F-theory diagrams such as (\ref{eq:su3flow}) capture already all the relevant information about the theory, using (\ref{eq:fi}), it is sometimes also convenient to summarize the theory in a different quiver notation:
\bea
\NNode{\vver{}{f_{1}}}{r_{1}}-\NNode{\vver{}{f_{2}}}{r_{2}}-\cdots-\NNode{\vver{}{f_{N-1}}}{r_{N-1}} \label{quivSU}
\eea
Here the blue node labelled $r_i$ denotes $SU(r_i)$ gauge group and the red node labelled $f_i$ denotes $SU(f_i)$ flavor symmetry. 

We can already see at this stage that the combinatorics of these quivers are related to two Young diagrams. Namely, if we consider the numbers
\begin{equation}\label{eq:si}
	\hat\rho_i= r_i-r_{i-1}\ ,
\end{equation}
we get a non-increasing sequence; dividing it into positive and negative regions, we naturally obtain the rows of two Young diagrams. For example, to the quiver 
\bea
\NNode{\vver{}{1}}{4}-\NNode{\vver{}{2}}{7}-\Node{}{8}-\Node{}{9}-\NNode{\vver{}{1}}{10}-\Node{}{10}-\Node{}{10}-\Node{}{10}-\Node{}{10}-\Node{}{10}-\NNode{\vver{}{1}}{10}-\Node{}{9}-\NNode{\vver{}{1}}{8}-\Node{}{6}-\Node{}{4}-\Node{}{2}
\label{quivex}
\eea
one associates the Young diagrams 
\begin{equation}\label{eq:young-ex}
	\rho_{\rm L}={\tiny \yng(4,3,1,1,1)} \qquad \rho_{\rm R}={\tiny \yng(2,2,2,2,1,1)}\ .
\end{equation}
One usually associates to these two partitions by reading the columns: in this case, $\rho^{\rm L}=[5,2^2,1]$ and $\rho^{\rm R}=[6,4]$ respectively.

\begin{figure}[ht]
\centering	
	\subfigure[\label{fig:D8-in}]{\includegraphics[width=8cm]{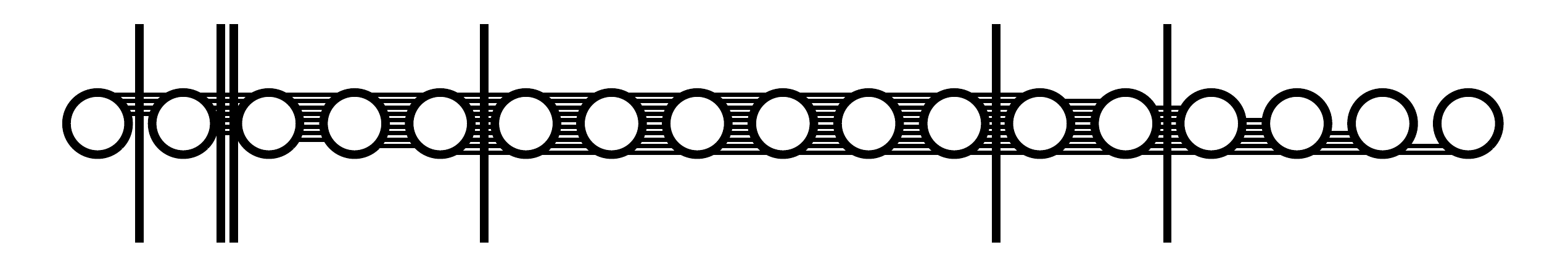}}
	\subfigure[\label{fig:D8-out}]{\includegraphics[width=8cm]{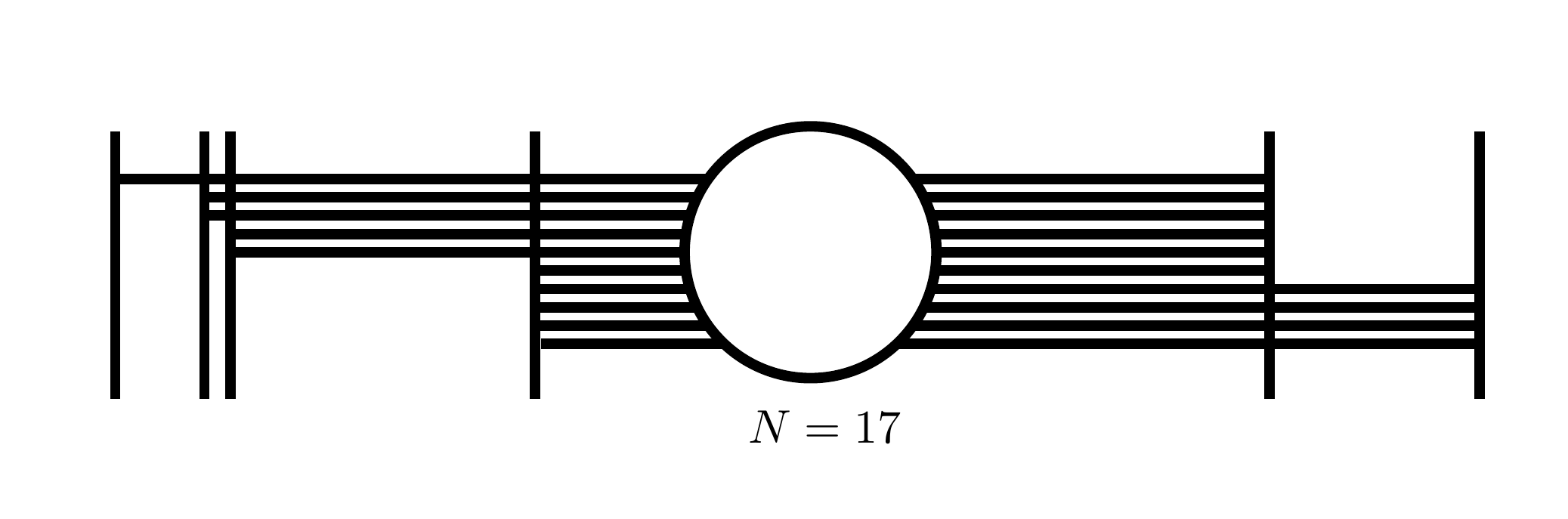}}
	\caption{\small Two equivalent brane diagrams engineering the theory in \eqref{quivex}. The nodes represent NS5-branes (spanning directions 0--5), the horizontal lines D6-branes (spanning 0--6), the vertical lines D8-branes (transverse to direction 6).}
	\label{fig:suk}
\end{figure}

In order to understand the Higgs moduli space of these theories, it is helpful to also keep in mind the IIA brane realization  \cite{Hanany:1997gh, Brunner:1997gk}  of these theories. This was also summarized in \cite[Sec.~2]{Cremonesi:2015bld}. For example, (\ref{quivex}) is realized by the brane diagrams in figure \ref{fig:suk}. The Young diagrams (\ref{eq:young-ex}) encode how many D8s intersect the various D6 segments in figure \ref{fig:D8-in}, and how many D6s end on the various D8s in figure \ref{fig:D8-out}. 

The Higgs moduli space dimension $d_H$ can now be computed as follows:
\bea \label{dHSU}
d_H &= \sum_{i=1}^{N-1} r_i f_i + \sum_{i=1}^{N-1} r_i r_{i+1} - \sum_{i=1}^{N-1} (r_i^2-1)~, \quad\text{with}~ r_0 = r_{N} =0 \nn\\
&= \frac{1}{2}\left( \sum_{i=1}^{N-1} f_i r_i \right)+{N-1} + \frac{1}{2} \sum_{i=1}^{N-1} r_i (-2r_i + r_{i-1} + r_{i+1} + f_i) \nn \\
&=  (N-1) + \frac{1}{2}  \sum_{i=1}^{N-1} f_i r_i~,
\eea
where we have used (\ref{eq:fi}). 

\paragraph{NS5 movement.}
Let us understand the two terms in (\ref{dHSU}) separately. Consider first the term $N-1$. This appears because the gauge groups are $SU(r_i)$ rather than $U(r_i)$; when quiver diagrams such as (\ref{quivSU}) are considered in lower dimensions, this is not so, and this summand does not appear. Its meaning is transparent in the IIA realization: it corresponds to the possibility of moving each of the $N-1$ NS5-branes in the directions transverse to the D6-branes. This possibility exists only in six dimensions, because we include in the theory the degrees of freedom of the NS5s (the tensor multiplets), which for analogous constructions in lower dimensions (such as D3--D5--NS5 diagrams \cite{hanany-witten}) would be too heavy. 

We can see this more precisely at the level of equations. Let us decompose the scalars in the bifundamental hypermultiplet between the $i$-th and $(i+1)-th$ gauge groups into two complex scalars $R_i$, $L_{i+1}$, and the fundamental hypers by $U_i$, $D_i$ (the letter describing the direction of the arrow, the index the gauge group which the arrow is leaving). Let us parametrize then the moduli space by the ``mesons'' $L_{i+1}R_i$, $R_i L_{i+1}$. 

Let us first see what happens in the original unHiggsed orbifold theory, which describes $N$ M5s on a $\bR^4/\bZ_k \times \bR$ singularity, or $N$ NS5s intersecting $k$ D6-branes. Here the gauge groups are all equal to $SU(k)$. The F-term equations constrain the traceless part of $L_{i+1} R_i - R_{i-1} L_i$; thus they can be written as
\begin{equation}\label{eq:F}
	L_{i+1}R_i = R_{i-1} L_i + \nu_i 1_{r_i}\ 
\end{equation}
for some $\nu_i$. The last term would not be present had the gauge groups been $U(k)$. If one of the $\nu_i\neq 0$, we see that there is no solution where the gauge group is completely unbroken: the product of the $i$-th and $(i+1)$-th copy of $SU(k)$ is broken to its diagonal. 

Let us also consider an example where the gauge groups are linearly increasing, such as the case $r_i= i n_0$: so the gauge group is $SU(n_0) \times SU(2 n_0)\times \ldots$. This is realized in IIA by NS5-branes in a region with Romans mass $F_0= \frac{n_0}{2\pi l_s}$; the Bianchi identity for $F_2$ requires that $n_0$ D6s end on each NS5. We again have no $U_i$, $D_i$ and the F-term equations read as in (\ref{eq:F}). A useful fact is that, if we have an $n\times m $ matrix $A$ and an $m \times n$ matrix $B$, with $n>m$, the characteristic polynomial $P_\lambda(B A)$ is equal to $P_\lambda(A B) \lambda^{n-m}$. Thus the eigenvalues of $BA$ are the same of those of $AB$, plus 0 repeated $n-m$ times. 
 
Let us consider what happens when 
\begin{equation}\label{eq:nui0}
	\nu_i=0\ .
\end{equation}
In this case, the first F-term equation reads $L_2 R_1 =0$. Then it follows that $R_1 L_2$ has zero eigenvalues; in other words, it is nilpotent. (We start seeing here the role of nilpotent matrices, which we will soon explore more.) The next F-term equation reads $R_1 L_2=L_3 R_2$. It follows that $R_2 L_3$ also has zero eigenvalues, and so on. So all the mesons have zero eigenvalues: they are nilpotent.

Let us now suppose $\nu_1\neq 0$, while $\nu_2 = \nu_3 =\ldots =0$. Now $L_2 R_1 = \nu_1 1_{n_0}$. It follows this time that $R_1 L_2$ has eigenvalues $\nu_1$ and $0$, both repeated $n_0$ times. Continuing as before, we see that all subsequent mesons $R_i L_{i+1}$ have $\nu_1$ eigenvalue repeated $n_0$ times, while the rest  of the eigenvalues are 0. This block of eigenvalues equal to $\nu_1$ indicates that if we move an NS5 in the directions transverse to the D6s, it must carry with it $n_0$ D6s, again because of the Bianchi identity. Thus in this case not only does the gauge group get broken, but the breaking ``propagates'' down the chain. 

Thus we see that the $N-1$ complex numbers $\nu_i$ represent transverse motions of the NS5s. We have analyzed the F-terms; a similar analysis would apply to D-terms, where $N-1$ real numbers $\lambda_i$ would appear, with a similar role. Together, the $\nu_i$ and $\lambda_i$ are the motions of the NS5s in the three dimensions transverse to them. They are also to be considered as the hyper-momentum map of $N-1$ hypermultiplets.\footnote{In the case without Romans mass, when a lift to M-theory is possible, the NS5s lift to M5s, which have four transverse directions which can be more directly identified with the hypermultiplets. In projecting to IIA, one loses an $S^1$ and ends up with only three transverse directions.}

 In the case where all the $\nu_i=0$, the mesons are nilpotent. The second term in (\ref{dHSU}) parameterizes these nilpotent mesons: we now turn to it.

\paragraph{The nilpotent part.}
Having discussed a bit the meaning of the summand $N-1$ in (\ref{dHSU}), we now come to the main part, the summand $\frac12 \sum_i f_i r_i$. This term is the dimension of the Higgs moduli space for the three-dimensional theory associated to a linear quiver such as (\ref{quivSU}). 

Such three-dimensional theories have been studied extensively, and some of the results carry over to the present situation. The brane engineering is very similar to that in figure \ref{fig:suk}, except that we have D3- and D5-branes instead of D6- and D8-branes. In \cite{Gaiotto:2008sa} the Higgs moduli space was interpreted in terms of the Nahm equations describing the transverse excitations of the D3s. In order to arrive at this interpretation, one needs to pull all the D5-branes on one side of the diagram, and all the NS5-branes on the other side. 

We can do the same for our NS5--D6--D8 diagram, pulling all the NS5 on one side. The number of D6-branes ending on each D8 can now be summarized in a partition $\sigma$, just as how the D6-branes ending on the D8s in figure \ref{fig:suk} for the theory (\ref{quivex}) is captured by Young diagrams in (\ref{eq:young-ex}). Similarly, the information of how many D6s end on each NS5 can be summarized in a partition $\rho$. These partitions can be easily related to the ones in figure \ref{fig:suk}. Let us give names to the transverse partitions $\rho^t_{\rm L}= [\hat\rho_1^{\rm L},\ldots,\hat\rho_L^{\rm L}]$ and similarly $\rho_{\rm R}^t= [\hat\rho_1^{\rm R},\ldots,\hat\rho_R^{\rm R}]$; in other words, let us consider the partitions obtained by reading the lengths of the rows of the Young diagrams (\ref{eq:young-ex}), rather than of the columns. Then we have 
\begin{align}\label{eq:rs}
	 &\sigma^t=[\hat\rho_1^{\rm L}+\hat\rho_1^{\rm R},\hat\rho_2^{\rm L}+\hat\rho_1^{\rm R},\ldots,\hat\rho_L^{\rm L}+\hat\rho_1^{\rm R},\underbrace{\hat\rho_1^{\rm R},\ldots,\hat\rho_1^{\rm R}}_{N-L-R},- \hat\rho_R^{\rm R}+\hat\rho_1^{\rm R},- \hat\rho_R^{\rm R-1}+\hat\rho_1^{\rm R},\ldots,-\hat\rho_2^{\rm R}+ \hat\rho_1^{\rm R}] \ \nonumber,\\ 
	&\rho= [(\hat\rho_1^{\rm R})^N] .
\end{align}
(Notice that $\rho$ is a rectangular Young diagram because of the stringent anomaly cancellation conditions for six-dimensional theories; this would not be so for three-dimensional theories.) While $\rho_{\rm L}$ and $\rho_{\rm R}$ have $k$ boxes, the number of boxes of $\rho$ and $\sigma$ is the larger number $K\equiv N \hat\rho_1^{\rm R}$. Recall that the number of gauge groups is $N-1$, while $N$ is the number of NS5-branes engineering the theory (see figure \ref{fig:suk}).  
The formula (\ref{eq:rs}), and the procedure to obtain it, is shown graphically in \cite[Fig.~7]{DelZotto:2014hpa}. As an example, for the two partitions in (\ref{eq:young-ex}), relative to the quiver in (\ref{quivex}), we get the partitions $\sigma^t=[6,5,3^3,2^6,1^2]$; $\rho=[2^{17}]$. 

The original result in \cite{Gaiotto:2008sa} was a description of the Higgs moduli space in terms as an intersection
\begin{equation}\label{eq:OS}
	{\cal O}_{\rho^t} \cap S_\sigma
\end{equation}
between the nilpotent orbit ${\cal O}_{\rho^t}$ associated to $\rho$ and a space $S_\sigma$ called ``Slodowy slice,'' which intersects the orbit ${\cal O}_\sigma$ transversely in a single point. The nilpotent orbit ${\cal O}_\rho$ is simply defined as the space of all elements in $SU(K)$ which are conjugate to a nilpotent matrix $N_\rho$, defined as a block-diagonal matrix whose blocks are the Jordan matrices
\begin{equation}
	\left(\begin{array}{cccccc}
		0 & 1 & 0 &\cdots & & 0\\ 
		 \cdots & 0 & 1 & 0 & \cdots & 0\\
		  &   & \ddots & \ddots & &\vdots\\
		  &   & 	 & \ddots & &\vdots\\
		  &   &   & \cdots& 0 & 1 \\
		  &   &   &  & \cdots & 0 
	\end{array}\right)
\end{equation} 
whose dimensions are $\rho_i$, the lengths of the columns of $\rho$. The quaternionic dimension of the corresponding orbit ${\cal O}_\rho$ in $SU(K)$ is given by 
\begin{equation}\label{eq:dimOrho}
	d_{{\cal O}_\rho}  = \frac12 (K^2 - \sum \hat\rho_i^2)\ ,
\end{equation}
where the $\hat\rho_i$ are the lengths of the rows of $\rho$, or in other words $\rho^t=[\hat\rho_1,\hat\rho_2,\ldots]$ as in our earlier notation in (\ref{eq:rs}).  One can also associate to ${\cal O}_\rho$ a subalgebra $\mathfrak{su}(2) \subset \mathfrak{su}(K)$, namely three matrices $\ell_i$ in the Lie algebra of $\mathfrak{su}(K)$ such that $[\ell_i,\ell_j]= \epsilon_{ijk}\ell_k$ such that $N_\rho=\ell^1 + i \ell^2$.  The Slodowy slice $S_\rho$ is defined as the space of elements which are conjugate to $N_\rho + x^\alpha v_\alpha$, where $v_\alpha$ are such that $[\ell_1-i\ell_2, v_\alpha]=0$. The intersection of $S_\rho$ with ${\cal O}_\rho$ is simply given by the point $\{x^\alpha=0\}$; the condition on the $v_\alpha$ is such that the intersection is transversal.   

One can check (\ref{eq:OS}) at the level of the dimension; using (\ref{eq:dimOrho}) one can show
\begin{equation}
	\frac12 \sum_i f_i r_i = d_{{\cal O}_\sigma}- d_{{\cal O}_\rho}\ .
\end{equation}
(This is also true for the more general three-dimensional quivers, for which $\rho$ may not be rectangular as in (\ref{eq:rs}).) 

All this already shows that nilpotent orbits play an important role for the moduli spaces of the quivers \ref{quivSU}. However, $\rho$ and $\sigma$ are not associated to orbits of our $G=SU(k)$; they are partitions of the larger number $K= N \hat\rho^{\rm R}_1$. It is natural to wonder if one can get a description directly in terms of the partitions $\rho_{\rm L}$, $\rho_{\rm R}$ of $k$. For this:
\begin{align}\label{eq:OLR}
\begin{split}
	d_{{\cal O}_{\rho^t}}- d_{{\cal O}_\sigma}&= \frac12\sum_i (\rho_i^2- \hat\sigma_i^2)\\
	&=\frac12\left[(\hat \rho_1^{\rm L}+\hat \rho^{\rm R}_1)^2+
		(\hat \rho_2^{\rm L}+\hat \rho^{\rm R}_1)^2 +\ldots+(\hat \rho_1^{\rm R})^2
		+\ldots +(-\hat \rho_2^{\rm R}+\hat \rho^{\rm R}_1)^2
		\right.\\
	 &\left. \hspace{2cm}- (\hat\rho^{\rm R}_1)^2 -(\hat\rho^{\rm R}_1)^2 -\ldots -(\hat\rho^{\rm R}_1)^2 -\ldots-(\hat\rho^{\rm R}_1)^2  \right]\\
	&=\frac12\sum_i (\hat\rho^{\rm L}_i)^2 + \frac12\sum_i (\hat\rho^{\rm R}_i)^2 +  \hat \rho^{\rm R}_1 \left(\sum_i \hat \rho^{\rm L}_i - \sum_i \hat \rho^{\rm R}_i\right)\\
	&=\frac12\sum_i
	 ((\hat\rho^{\rm L}_i)^2+(\hat\rho^{\rm R}_i)^2)= 
	k^2 - d_{{\cal O}_{\rm L}}- d_{{\cal O}_{\rm R}}\ .	
\end{split}	
\end{align}
In the first step we have used (\ref{eq:dimOrho}). In the second, we have used (\ref{eq:rs}), with the $\rho_i^2$ on one line and the $\hat \sigma_i$ (the lengths of rows of $\sigma$) on the next. In the third step we have simplified each square with the square below it. In the fourth step we have used the fact that $\sum_i \hat\rho^{\rm L}_i = \sum_i \hat\rho^{\rm R}_i(=k)$. Finally, in the fifth step we have used (\ref{eq:dimOrho}) again, this time for the orbits ${\cal O}_{\rho^{\rm L}}\equiv {\cal O}_{\rm L}$ and ${\cal O}_{\rho^{\rm R}}\equiv {\cal O}_{\rm R}$ of $SU(k)$.

Our result in (\ref{eq:OLR}) strongly suggests that this part of the moduli space, of dimension $\frac12 \sum_i f_i r_i$, can also be viewed as an intersection of two Slodowy slices for orbits in $SU(k)$, transverse to the nilpotent orbits ${\cal O}_{\rm L}$ and ${\cal O}_{\rm R}$. 

We can examine a couple of easy examples. Let us first take  $\rho_{\rm L}=\rho_{\rm R}=[1^k]$, so that ${\cal O}_{\rm L}={\cal O}_{\rm R}$ are both the zero orbit. This corresponds to the quiver whose gauge groups are all $SU(k)$: the unHiggsed theory of M5-branes at a $\bR^4/\bZ_k\times \bR$ singularity.
In this case (\ref{eq:OLR}) gives $k^2$. Indeed the Slodowy slices in this case are all $SU(k)$. If we look explicitly at the F-term equations, the first reads $U_1 D_1 = L_2 R_1$; $U_1$ and $D_1$ are $k\times k$ matrices, so there are no restrictions on the eigenvalues and rank of the first meson $L_2R_1$. The subsequent F-term equations only say that this is equal to $R_1 L_2=L_3 R_2$ and so on, so they also put no restriction on the mesons. 

For another example, $\rho_{\rm L}= [k]$, $\rho_{\rm R}=[1^k]$; now ${\cal O}_{\rm R}$ is again the zero orbit, but ${\cal O}_{\rm L}$ is the largest nilpotent orbit, with complex dimension $k^2-k$. Now (\ref{eq:OLR}) gives $k$. In this case, $r_i=i$ for $i\le k$, $r_i =k$ for $i\ge k$; we have $f_k=1$ and all the other $f_i=0$. The first $k-1$ F-term equations impose the first mesons $L_{i+1} R_i$, $i\le k-1$ to be nilpotent, because of the argument discussed around (\ref{eq:nui0}). The $k$-th F-term equation reads $L_{k+1}R_k= R_{k-1}L_k + U_k D_k$. Since $f_k=1$, $U_k D_k$ is a rank 1 matrix; so $L_{k+1}R_k$ is the sum of a rank 1 matrix and of a nilpotent one. By a gauge transformation one can put such a matrix in the form of a Slodowy slice, or in the perhaps more familiar form 
\begin{equation}
	\left(\begin{array}{cccccc}
		0 & 1 & 0 &\cdots & & 0\\ 
		 \cdots & 0 & 1 & 0 & \cdots & 0\\
		  &   & \ddots & \ddots & &\vdots\\
		  &   & 	 & \ddots & &\vdots\\
		  &   &   & \cdots& 0 & 1 \\
		a_k & a_{k-1}  & \cdots  &  & \cdots & a_1 
	\end{array}\right)\ .
\end{equation}
This shows again the relation to Slodowy slices.

\bigskip 

Let us now summarize this subsection. The Higgs moduli space for linear chains of $SU(r_i)$ gauge groups such as (\ref{quivSU}) has dimension (\ref{dHSU}). The term $N-1$ is peculiar to our six-dimensional theories, and in the string-theoretic engineering it represents motion of the NS5s away from the D6s. The term $\frac12 \sum_i f_i r_i$ can be rewritten in terms of nilpotent orbits as in (\ref{eq:OLR}). Putting all together, we arrive at 
\begin{equation}\label{eq:dimsuk}
	d_H(\CO_{\rm L},\CO_{\rm R}) = N-1 + k^2 - d_{{\cal O}_{\rm L}} - d_{{\cal O}_{\rm R}} \ ,
\end{equation}
which agrees with (\ref{eq:dim}) for $G=SU(k)$.

In the rest of the paper, we will see that (\ref{eq:dim}) is true for other $G$'s.

\subsection{Anomalies}\label{sub:anom}

There has recently been great progress in the understanding of anomalies and their behavior under RG flows between 6D SCFTs \cite{Intriligator:1997dh, Ohmori:2014pca, Ohmori:2014kda, Cordova:2015vwa, Heckman:2015ola, Cordova:2015fha, Beccaria:2015ypa, Fei:2015oha, Heckman:2015axa, Shimizu:2016lbw}.  The algorithm of \cite{Ohmori:2014kda} enabled the computation of the anomaly polynomial
\begin{equation}
I = \alpha c_2(R)^2 + \beta c_2(R) p_1(T) + \gamma p_1(T)^2 + \delta p_2(T)
\end{equation}
for any 6D SCFT.  Subsequently, \cite{Cordova:2015fha} showed that the coefficients $\alpha, \beta, \gamma$ and $\delta$ are related to the coefficient $a$ of the Euler density in the trace anomaly via
\begin{equation}
a = \frac{16}{7}(\alpha-\beta+\gamma) + \frac{6}{7} \delta.
\label{eq:a}
\end{equation}
(\cite{Beccaria:2015ypa} gave similar relations for the coefficients $c_i$.)
\cite{Cordova:2015fha} also proved that $a$ decreases along supersymmetric RG flows triggered by giving vevs to scalar fields in tensor multiplets (thereby moving out on the ``tensor branch" of these theories).  Using these results, \cite{Heckman:2015axa} performed a vast sweep of flows triggered by giving vevs to fundamental hypermultiplets (thereby moving out on the ``Higgs branch" of these theories), yielding strong evidence that $a$ decreases along these flows as well.  Nonetheless, a rigorous proof that $a$ decreases along Higgs branch flows is still lacking.  In the following sections, we will derive explicit formulae for the anomaly coefficients of the T-brane theories and demonstrate that the $a$-theorem is satisfied for the flows.

The coefficients $\alpha$ and $\beta$ change between the UV and IR theories of Higgs branch flow, and this change is absorbed by Nambu-Goldstone (NG) bosons \cite{Cordova:2015fha}. Poincar\'e symmetry is preserved under Higgs branch flows, and as a result one can apply the 't Hooft anomaly matching condition to argue that $\Delta\delta$ and $\Delta \gamma$ vanish along the flows.  $\gamma$ and $\delta$ do not match for the interacting part of the SCFTs in the UV and IR, but free hypermultiplets appear in the IR of such a flow to compensate for the difference in these anomaly coefficients.  These hypermultiplets are simply the ones that have not been eaten and thus parametrize the Higgs moduli space. In a flow between two theories without vector multiplets, this means that $\delta$ should be related to the quaternionic dimension of the moduli space $d_H$ by $\delta=-\frac1{24\cdot 60} d_H$. In our case, where not all the gauge groups are Higgsed, we can still conclude
\begin{equation}\label{eq:DdeltaDdH}
	\Delta \delta =- \frac 1{24\cdot 60} \Delta d_H\ .
\end{equation}
Note that $\delta$ always increases along an RG flow, and correspondingly $d_H$ decreases.

\section{Universal formulae for anomaly coefficients} \label{sec:anomaly}
In this section, we present universal formulae for computing the anomaly coefficients $\gamma$ and $\delta$ that are valid for all T-brane theories and discuss the relation to the Higgs moduli space dimension of the theories, thereby deriving \eref{eq:dim}.  We also present a formula that captures the universal behavior of the coefficient $\beta$, though it also contains non-universal terms that depend on the specific T-brane theory in question. 

\subsection{The coefficient $\delta$} \label{sec:coeffdelta}
The anomaly coefficient $\delta$ is given by
\bea
\delta= -\frac{1}{24 \cdot 60}\left( 29 n_T + n_H - n_V\right)~, \label{deltaquiv}
\eea
where $n_T$ is the number of tensor multiplets in the F-theory quiver, $n_H$ denotes the total degrees of freedom of the hypermultiplets counting in the quaternionic unit, and $n_V$ is the number of the vector multiplets which is equal to the total dimension of the gauge groups.  

\paragraph{The relation with the dimension of the orbit.} The coefficient $\delta$ can also be computed recursively.  Let $T_{\CO_L, \CO_R }$ and $T_{\CO_L', \CO_R'}$ be theories associated with pairs of nilpotent orbits $(\CO_L, \CO_R)$ and $(\CO_L',\CO_R')$, whose quaternionic dimensions are $d_{\CO_L}$, $d_{\CO_R}$ and $d_{\CO_L'}$, $d_{\CO_R'}$ respectively. Suppose the anomaly coefficients $\delta$ of the two theories are $\delta_{\CO_L, \CO_R}$ and $\delta_{\CO_L',\CO_R'}$, respectively.  We find an interesting relation between
\begin{equation}
	\Delta \delta\equiv \delta_{\CO_L, \CO_R}- \delta_{\CO_L',\CO_R'} \quad \text{and} \quad \Delta d \equiv (d_{\CO_L}+ d_{\CO_R})- (d_{\CO_L'} + d_{\CO_R'})\ ,
\end{equation}
 as follows:
\bea
\Delta \delta= \frac{1}{24 \cdot 60} \Delta d = -\frac{1}{24 \cdot 60} (29 \Delta n_T + \Delta n _H -\Delta n _V)~, \label{diffdelta}
\eea
where $\Delta n_T$, $\Delta n_H$ and $\Delta n_V$ are differences between the numbers of tensor multiplets, hypermultiplets and vector multiplets in theory $T_{\CO_L, \CO_R }$ and theory $T_{\CO_L', \CO_R'}$.   Suppose $\CO_L$ is below $\CO_L'$ with respect to the the arrow in the Hasse diagram (\ie~ the theory corresponding to $\CO_L'$ flows to the one corresponding to $\CO_L$).  It follows that $d_{\CO_L} > d_{\CO_L'}$ and so $\delta_{\CO_L,\CO_R} > \delta_{\CO_L',
\CO_R'}$.  Thus, $\delta$ increases along the flow in the nilpotent hierarchy, as expected.  We demonstrate \eref{diffdelta} in examples below.

Importantly, $\delta$ can also be used as a proxy for the Higgs moduli space dimension.  We can now derive (\ref{eq:dim}). In \eref{diffdelta}, let us set ${\CO_L'} = {\CO_R'}=0$ to be the trivial orbit of group $G$.  The corresponding theory $T_{0,0}$ is simply the worldvolume theory on multiple M5-branes on $\BC^2/\Gamma$, where $\Gamma$ is the discrete group related to $G$ by McKay correspondence. 
Using \eref{diffdelta} and (\ref{eq:DdeltaDdH}) we obtain \eref{eq:dim}:
\bea
d_H(\CO_{\rm L},\CO_{\rm R}) - d_{H}(\vec{0})  = d_{{\cal O}_{\rm L}} + d_{{\cal O}_{\rm R}}\ .
\eea
Notice that using \eref{eq:DdeltaDdH} and \eref{diffdelta}, we see that the combination $29 \Delta n_T + \Delta n _H -\Delta n _V$ in the right hand side of \eref{diffdelta} is equal to the difference $\Delta d_H$ between the Higgs moduli space dimension of theory $T_{\CO_L, \CO_R }$ and that of theory $T_{\CO_L', \CO_R'}$.

The second equality of \eref{diffdelta} suggests the existence of a conserved quantity along the flow from one orbit to another.  For a pair of orbits $\CO_L, \CO_R$ of the group $G$, we find that the following relations hold:
\bea 
d_{\CO_L} + d_{\CO_R} + 29 n_T+ n_H -n_V  =  30\mathfrak{n}+\dim(G) +1~. \label{dimTHG}
\eea
where $n_T$, $n_H$ and $n_V$ are the numbers of tensor multiplets, hypermultiplets and vector multiplets in the quiver associated with the orbits $\CO_L$, $\CO_R$; $\dim(G)$ is the dimension of the group $G$; and $\mathfrak{n}$ is the number of $-2$ curves in the F-theory quiver after blowing down all $-1$ curves \cite{Heckman:2013pva}.  We discuss how to compute $\mathfrak{n}$ in examples below.  

The right hand side of \eref{dimTHG} can be derived by inspecting various nilpotent orbits.  The $\mathfrak{n}$-independent part of \eref{dimTHG} can be obtained by considering the conformal matter theory of group $G$ (see Table \ref{tab:confmatter}) where $\mathfrak{n}=0$ and $d_{\CO}=0$.  The coefficient of $\mathfrak{n}$ can be fixed by considering a long chain of F-theory quiver, such as the theories on $N$ M5-branes on $\BC^2/\Gamma_G$.  In this specific case, $\mathfrak{n}=N-1$ is equal to the number of times that gauge group $G$, in the McKay correspondence to $\Gamma_G$, appears in the quiver (see Example 1 below).   The quantity on the right hand side of \eref{dimTHG}, and hence $\mathfrak{n}$, is constant along the flow from one orbit to another.\footnote{It is always possible to write down  the F-theory quivers that are sufficiently long so that $\mathfrak{n}$ is constant along the flow in the nilpotent hierarchy.}

\begin{table}[h]
\begin{center}
\begin{tabular}{|c|c|c|c|c|c|}
\hline
Group & Conformal matter quiver & $n_T$ & $n_H$ & $n_V$   \\
\hline
$SO(2k)$  & {\footnotesize$[SO(2k)]   \,\, {\overset{\mathfrak{usp}(2k-8)}1}  \,\,  [SO(2k)]$} &  1 & {\footnotesize $2k(2k-8)$} & {\footnotesize $(k-4)(2k-7)$} \\
$E_6$ & {\footnotesize $[E_6]  \,\, 1 \,\, {\overset{\mathfrak{su_3}}3}  \,\,   1 \,\,[E_6]$} & 3 & 0 & 8    \\
$E_7$ & {\footnotesize $[E_7] \,\, 1\,\,  \overset{\mathfrak{su_2}}2 \,\, \overset{\mathfrak{so_7}}3  \,\,  \overset{\mathfrak{su_2}}2\,\, 1 \,\,  [E_7]$} & 5& 16 & 27 \\
$E_8$ & {\footnotesize $[E_8] \,\, 1\,\, 2 \,\, \overset{\mathfrak{su_2}}2 \,\, \overset{\mathfrak{g_{2}}}3  \,\, 1 \,\, \overset{\mathfrak{f_{4}}}5 \,\,  1 \,\, \overset{\mathfrak{g_{2}}}3 \,\, \overset{\mathfrak{su_2}}2 \,\, 2 \,\, 1\,\, [E_8]$} &11&16  &86  \\
\hline
\end{tabular}
\end{center}
\caption{Various parameters for conformal matter theories.}
\label{tab:confmatter}
\end{table}%

From \eref{deltaquiv} and \eref{dimTHG}, we obtain another formula for the anomaly coefficient $\delta$ as follows:
\bea \label{deltaalt}
\delta= - \frac{1}{24 \cdot 60} \left[ 30\mathfrak{n} + \dim(G) +1- d_{\CO_L} - d_{\CO_R} \right]~.
\eea
Along an RG flow between successive theories, the first three terms in brackets remain constant; the only variation among $\delta$ between theories in an RG hierarchy comes from the variation in $d_{\CO_L} + d_{\CO_R}$.  Using \eref{eq:DdeltaDdH} and \eref{diffdelta}, this in turn gives us an expression for the change in the quaternionic dimension of the Higgs moduli space $d_H$ under a flow between theories labeled by nilpotent orbits:
\bea
d_H^{UV} -d_H^{IR} = (d_{\CO_L}^{IR} + d_{\CO_R}^{IR})- (d_{\CO_L}^{UV} + d_{\CO_R}^{UV})~.
\eea

Subsequently we apply the above formulae to several non-trivial examples.

\paragraph{Example 1.}  For the $G=SU(k)$ theories, reviewed in section \ref{sub:suk}, formula \eref{deltaquiv} reduces to
\bea
\delta = -\frac{1}{24 \cdot 60} \left[\frac{1}{2} \sum_{i=1}^{N-1} f_i r_i + 30 (N-1)  \right]~,
\eea
Here, $N-1=\mathfrak{n}$ is the number of tensor multiplets in the quiver \eref{quivSU}. This is in agreement with (3.9) of \cite{Cremonesi:2015bld}.

\paragraph{Example 2.} Let us compute $\delta$ for the worldvolume theories of $Q$ M5-branes on $\BC^2/\Gamma_G$ where $G= SO(2n),~ E_6, ~E_7,~E_8$.  The necessary information is given in Table \ref{tab:QM5}.
\begin{table}[h]
\begin{center}
{\footnotesize
\begin{tabular}{|c|c|c|c|c|c|c|c|c|c|}
\hline
$G$ & $n_H$ & $n_V$   & $n_T$ \\
\hline
$SO(2k)$    & {\footnotesize $2k(2k-8)Q$}  &\vtop{ \hbox{\footnotesize $(k-4)(2k-7)Q$} \hbox{\footnotesize $+k(2k-1)(Q-1)$} } & {\footnotesize $2Q-1$} \\
$E_6$  & 0   &  $8N+78(Q-1)$   & { $4Q-1$}   \\
$E_7$  & $16Q$   &  $27Q+133(Q-1)$ & { $6Q-1$}   \\
$E_8$  & $16Q$   &  $86Q+248(Q-1)$  & { $12Q-1$} \\
\hline
\end{tabular}}
\end{center}
\caption{Information about the worldvolume theories of $N$ M5-branes on $\BC^2/ \Gamma_G$, including the anomaly coefficient $\delta$.}
\label{tab:QM5}
\end{table}%

Using  \eref{deltaquiv}, we find that 
\bea
\delta &= -\frac{1}{24\cdot 60} \left[ 30(Q-1) + (\dim G)+1\right]~,
\eea
where $\dim G$ is the dimension of group $G$ in McKay correspondence with $\Gamma$. This is in agreement with \cite{Ohmori:2014kda}.   

Comparing this result to \eref{deltaalt}, with $d_{\CO}=0$, we see that $\mathfrak{n} =Q-1$; this is the number of times that the curve $\overset{\mathfrak{e_{8}}}{12}$ appears in the F-theory quiver.  

\paragraph{Example 3.} For the remainder of this subsection, we specialize to the case in which $\CO_R = \CO_R'$ is taken to be the trivial nilpotent orbit, and we label a theory $T_{\CO_L = \CO, \CO_R = 0}$ simply by $T_\CO$, with $\delta_{\CO} \equiv \delta_{\CO_L=\CO,\CO_R=0}$.

Let us take $\CO$ to be the minimal orbit.  For concreteness, we focus on the orbit $A_1$ of $E_8$, whose F-theory quiver is given by
\bea \label{A1ofE8}
[E_7] \,\, 1\,\,  \overset{\mathfrak{su_{2}}}2 \,\, \overset{\mathfrak{g_{2}}}3  \,\, 1 \,\, \overset{\mathfrak{f_{4}}}5 \,\,  1 \,\, \overset{\mathfrak{g_{2}}}3 \,\, \overset{\mathfrak{su_{2}}}2 \,\, 2 \,\, 1\,\, \overset{\mathfrak{e_{8}}}{12}  \,\, 1 \,\,  2 \,\, \overset{\mathfrak{su_{2}}}2 \,\, \overset{\mathfrak{g_{2}}}3  \,\, 1 \,\, \overset{\mathfrak{f_{4}}}5 \,\,  1 \,\, \overset{\mathfrak{g_{2}}}3 \,\, \overset{\mathfrak{su_{2}}}2 \,\, 2 \,\, 1 \; \ldots \; [E_8]
\eea 
As usual, let $\mathfrak{n}$ be the number of $-2$ curves after blowing down all $-1$ curves. Using \eref{deltaquiv} with $n_T = 10+12 \mathfrak{n} $, $n_H=16 (\mathfrak{n} +1)$ and $n_V = 86(\mathfrak{n}+1)+248(\mathfrak{n})$, we obtain
\bea
\delta = -\frac{1}{24 \cdot 60} [30\mathfrak{n}+220]= -\frac{1}{24 \cdot 60} [30\mathfrak{n}+248-(29-1)] ~,
\eea
It can also be checked that this is consistent with \eref{diffdelta}.  Let us take $\CO_L = 0$, $\CO'_L = A_1$ and $\CO_R = \CO'_R = 0$.   We find that
\bea
\Delta \delta &=- \frac{29}{24 \cdot 60}~, \nn \\
\Delta d  &=  0-29=-29~, \nn \\
\quad 29 \Delta n_T + \Delta n _H -\Delta n _V &= 29 \times 1 =29~. 
\eea
Thus the fact that the minimal orbit of $E_8$ has dimension 29 is reinterpreted as the contribution to $\delta$ of a single tensor multiplet. 

\noindent {\it General formula.} Using \eref{deltaalt}, we obtain the following general formula for the T-brane theory associated with the minimal nilpotent orbit of $G$, whose dimension is $d_{A_1}= h^\vee_G-1$,
\bea
\delta_{\text{min}} =  -\frac{1}{24 \cdot 60} [30\mathfrak{n}+ \dim(G) - h^\vee_G+2]~,
\eea
where $h^\vee_G$ is the dual Coxeter number of $G$.

\paragraph{Example 4.} Another instructive example consists of taking $\CO_R=0$ and taking $\CO_L=\CO$ to be the subregular and the principal orbits.  \\~\\
\noindent {\it The group $E_6$.} For concreteness, we first consider $E_6(a_1)$ and $E_6$ of the group $E_6$:
\bea 
E_6 (a_1):  \qquad {\overset{\mathfrak{su_{2}}}2}  \,\,   {\overset{\mathfrak{g_2}}3}  \,\, 1 \,\, \,  &\overset{\mathfrak{f_{4}}}5 \,\, \, \, 1 \,\,  {\overset{\mathfrak{su_{3}}}3}  \,\,   1 \,\,\overset{\mathfrak{e_{6}}}6  \,\, 1 \,\, {\overset{\mathfrak{su_{3}}}3}  \,\,   1 \,\, \ldots \, \, [E_6]  \label{quivE6a1} \\
E_6:  \hspace{2.3cm} & {2}   \,\, \,    {\overset{\mathfrak{su_{2}}}2}  \,\,   {\overset{\mathfrak{g_2}}3}  \,\, 1 \,\, \overset{\mathfrak{f_{4}}}5 \,\, 1 \,\,  {\overset{\mathfrak{su_{3}}}3}  \,\,   1 \,\, \ldots \, \, [E_6] \label{quivE6}~.
\eea
The above two lines are written to show that the empty curve $2$ in the theory corresponding to the orbit $E_6$ descends from the curve $\overset{\mathfrak{f_{4}}}5$ in the theory corresponding to the orbit $E_6(a_1)$, and so on.   

The anomaly coefficients $\delta$ for both theories can be computed using \eref{deltaquiv},
{\small
\bea
\begin{array}{lll}
-1440 \delta_{E_6(a_1)}(\mathfrak{n}) &= 29 [7+4(\mathfrak{n}-3)] + 8 - [77+ 86(\mathfrak{n}-3) ] &=  30 (\mathfrak{n}-4)+164~, \\
-1440 \delta_{E_6}(\mathfrak{n}) &= 29 [8+4(\mathfrak{n}-4)] + 8 - [77+ 86(\mathfrak{n}-4) ] &=  30 (\mathfrak{n}-4)+163~,
\end{array}
\eea}
It can also be checked that this is consistent with \eref{diffdelta}.  Let us take $\CO_L = E_6(a_1)$, $\CO'_L = E_6$ and $\CO_R = \CO'_R = 0$.   We find that
\bea
\Delta \delta &=- \frac{1}{24 \cdot 60}~, \nn \\
\Delta d  &=  35-36=-1~, \nn \\
\quad 29 \Delta n_T + \Delta n _H -\Delta n _V &= (29 \times 3) -8 -78=1~. 
\eea

Note that in these cases, $\mathfrak{n}$ is NOT equal to the number of times $n_{\rm max}$ that $\overset{\mathfrak{e}_6}6$ appears in the quiver.  In fact, they are related by $\mathfrak{n} = n_{\rm max} +3 $ for the orbit $E_6(a_1)$ and $\mathfrak{n} = n_{\rm max} +4 $ for the orbit $E_6$.  Let us now demonstrate how to obtain $\mathfrak{n} =4$ in the case of $E_6$ orbit when $n_{\rm max}=0$.  For the sake of brevity, we only write down the curves that appear in the F-theory.  Starting from \eref{quivE6} and blowing-down the $(-1)$-curves step-by-step, we obtain
\bea
22315131  \quad \rightarrow \quad 22231 \quad \rightarrow \quad 2222~,
\eea
where in each step we subtract 1 from each of the numbers on the left and right of each $(-1)$-curve, namely $\ldots M \, \, 1 \, \, N \ldots \rightarrow (M-1)(N-1)$.  Since at the end there are four $(-2)$-curves, we obtain $\mathfrak{n}=4$ as expected.
\\~\\
\noindent {\it The group $E_8$.} We compare $E_8(a_1)$ and $E_8$ orbits of the group $E_8$:
{\small
\bea
E_8(a_1): \qquad  {\overset{\mathfrak{su_{2}}}2} \,\, \overset{\mathfrak{g_{2}}}3  \,\, 1 \,\, \overset{\mathfrak{f_{4}}}5 \,\,  1 \,\, \overset{\mathfrak{g_{2}}}3 \,\, \overset{\mathfrak{su_{2}}}2 \,\, 2 \,\, 1\,\, \overset{\mathfrak{e_{8}}}{11}   \,\, 1\,\, &2 \,\, \overset{\mathfrak{su_{2}}}2 \,\, \overset{\mathfrak{g_{2}}}3  \,\, 1 \,\, \overset{\mathfrak{f_{4}}}5 \,\,  1 \,\, \overset{\mathfrak{g_{2}}}3 \,\, \overset{\mathfrak{su_{2}}}2 \,\, 2 \,\, 1 \,\, \overset{\mathfrak{e_{8}}}{12} \,\,  1\,\, 2 \,\, \overset{\mathfrak{su_{2}}}2 \,\, \overset{\mathfrak{g_{2}}}3  \,\, 1 \,\, \overset{\mathfrak{f_{4}}}5 \,\,  1 \,\, \overset{\mathfrak{g_{2}}}3 \,\, \overset{\mathfrak{su_{2}}}2 \,\, 2 \,\, 1 \, \,  ... [E_8] ~, \nn \\
E_8:  \hspace{5.6cm} & 2 \,\, \overset{\mathfrak{su_{2}}}2 \,\, \overset{\mathfrak{g_{2}}}3  \,\, 1 \,\, \overset{\mathfrak{f_{4}}}5 \,\,  1 \,\, \overset{\mathfrak{g_{2}}}3 \,\, \overset{\mathfrak{su_{2}}}2 \,\, 2 \,\, 1\,\, \overset{\mathfrak{e_{8}}}{11}  \,\,  1\,\, 2 \,\, \overset{\mathfrak{su_{2}}}2 \,\, \overset{\mathfrak{g_{2}}}3  \,\, 1 \,\, \overset{\mathfrak{f_{4}}}5 \,\,  1 \,\, \overset{\mathfrak{g_{2}}}3 \,\, \overset{\mathfrak{su_{2}}}2 \,\, 2 \,\, 1 \, \,  ... [E_8]~. \nn
\eea}
Using \eref{deltaquiv}, we obtain
\bea
\delta_{E_8(a_1)}(Q) =  - \frac{30 (\mathfrak{n}-5)+280}{24 \cdot 60}~, & \qquad \delta_{E_8}(Q) =  - \frac{30 (\mathfrak{n}-5)+279}{24 \cdot 60}~.
\eea
It can also be checked that this is consistent with \eref{diffdelta}.  Let us take $\CO_L = E_8(a_1)$, $\CO'_L = E_8$ and $\CO_R = \CO'_R = 0$.   We find that
{\small
\bea
\Delta \delta &=- \frac{1}{24 \cdot 60}~, \nn \\
\Delta d  &=  -1~,  \\
\quad 29 \Delta n_T + \Delta n _H -\Delta n _V &= (29 \times 11) +8 +8 -3 -14-52-14-3-248=1~. \nn
\eea}
Once again, note that $\mathfrak{n}$ is not equal to the number of times $n_{\rm max}$ that $\overset{\mathfrak{e}_8}{12}$ appears in the quiver.  Indeed, we have $\mathfrak{n} = n_{\rm max} +4 $ for the orbit $E_8(a_1)$ and $\mathfrak{n} = n_{\rm max} +5 $ for the orbit $E_8$. Let us demonstrate how to obtain $\mathfrak{n} =5$ in the case of $E_8$ orbit when $n_{\rm max}=0$.  Starting from the F-theory quiver of the $E_8$ orbit and blowing-down the $(-1)$-curves step-by-step, we obtain
{\small
\bea
 222322191223221  \,\, \rightarrow \,\, 222321712321  \,\, \rightarrow \,\, 222315131  \,\, \rightarrow \,\, 222231  \,\, \rightarrow \,\, 22222~,
\eea}
where, as before, in each step we perform the following action: $\ldots M \, \, 1 \, \, N \ldots \rightarrow (M-1)(N-1)$.  Since at the end there are five $(-2)$-curves, we obtain $\mathfrak{n}=5$ as expected.

\noindent {\it General formula.} Using \eref{deltaalt}, we obtain the following general formula for the T-brane theory associated with the subregular and the principal orbit orbits of $G$, whose dimensions are respectively 
\bea \label{dimsubregprinc}
d_{\text{subreg}} &= \frac{1}{2} [\dim (G)- \mathrm{rank}(G))] -1~, \nn \\
d_{\text{princ}} &= \frac{1}{2} [\dim (G)- \mathrm{rank}(G)]~.
\eea
and hence
\bea
\delta_{\text{subreg}} &=  -\frac{1}{24 \cdot 60} \left[30\mathfrak{n} + \frac{1}{2} \dim(G) +\frac{1}{2} \mathrm{rank}(G)+2 \right]~, \\
\delta_{\text{princ}} &=  -\frac{1}{24 \cdot 60} \left[30\mathfrak{n}+ \frac{1}{2} \dim(G) +\frac{1}{2} \mathrm{rank}(G)+1 \right]~.
\eea

\paragraph{Example 5.} We now demonstrate \eref{diffdelta} using various orbits $E_8$ whose F-theory quivers exhibit certain interesting features.  We set $\CO_R = \CO'_R = 0$ and consider the following pairs of $(\CO_L, \CO'_L)$.
\bi
\item  $(\CO_L, \CO'_L) = (2A_2+A_1, 2A_2+2A_1)$.   The corresponding F-theory quivers are
\bea
2A_2+A_1: &  \qquad   [G_2] \,\, 1 \,\, \underset{[USp(2)]}{\overset{\mathfrak{f_{4}}}4} \,\,  1 \,\, \overset{\mathfrak{g_{2}}}3 \,\, \overset{\mathfrak{su_{2}}}2 \,\, 2 \,\, 1\,\, \overset{\mathfrak{e_{8}}}{12}  \,\, 1 \,\,  ... [E_8] \\
2A_2+2A_1: & \qquad  [USp(4)]   \,\,  {\overset{\mathfrak{f}_4}3} \,\,  1 \,\, \overset{\mathfrak{g_{2}}}3 \,\, \overset{\mathfrak{su_{2}}}2 \,\, 2 \,\, 1\,\, \overset{\mathfrak{e_{8}}}{12}  \,\, 1 \,\,  ... [E_8]
\eea
The $\mathfrak{f_4}$ gauge group has respectively one and two fundamental $\mathbf{26}$ flavor representations for he upper and lower theory (on which an $USp(2)$ and $USp(4)$ are acting, respectively). Thus in the flow we lose a tensor, but we gain 26 hypers. This matches the orbit difference:
\bea
\Delta d  &=  81- 84 = -3 ~,  \\
\quad 29 \Delta n_T + \Delta n _H -\Delta n _V &= 29+(26 - 2 \times 26) -0 =3 ~. \nn
\eea

\item  $(\CO_L, \CO'_L) = (A_4+A_2, A_5)$. This example illustrates a case where a ramified quiver appears after the flow. The corresponding F-theory quivers are
\bea
A_4 + A_2: & \qquad [SO(4)] \,\, \overset{\mathfrak{su}_2}2\,\, {\overset{\mathfrak{su}_2}2 } \,\,  \underset{[N_f=1]}{\overset{\mathfrak{su}_2}2} \,\, \overset{\mathfrak{su}_1}2 \,\,  1\,\, \overset{\mathfrak{e_{8}}}{12}  \,\, 1 \,\,  ... [E_8] \\
A_5: & \qquad [G_2] \,\, {\overset{\mathfrak{su}_2}2} \,\, 2 \,\,  1\,\, \underset{[SU(2)]}{\underset{2}{\underset{1}{\overset{\mathfrak{e_{8}}}{12}}}}  \,\, 1 \,\,  ... [E_8]~. \label{quivA5}
\eea
We find that
\bea
\Delta d  &=  97-98 = -1 ~,  \\
\quad 29 \Delta n_T + \Delta n _H -\Delta n _V &= 0+ (4+4 + 4+2+2 -8-1)- (3+3) = 1 ~, \nn
\eea
where the single hypermultiplet, denoted by $1$ in the second bracket in the last line, comes from the curves $2 \,\, [SU(2)]$ in \eref{quivA5}.
\item  $(\CO_L, \CO'_L) = (E_6(a_3), D_5)$. The corresponding F-theory quivers are
\bea
E_6(a_3): & \qquad[G_2] \,\, {\overset{\mathfrak{su}_2}2 } \,\,  2  \,\,  1\,\, \overset{\mathfrak{e_{8}}}{10}  \,\, 1 \,\, 2 \,\,  \overset{\mathfrak{su_{2}}}2 \,\, \overset{\mathfrak{g_{2}}}3  \,\, 1 \,\, \overset{\mathfrak{f_{4}}}5 \,\,  1  ... [E_8] \label{quivE6a3}\\ 
D_5: & \qquad  [SO(7)] \,\,   \overset{\mathfrak{su_{2}}}2 \,\,  1\,\, \overset{\mathfrak{e_{7}}}8  \,\, 1\,\,  \overset{\mathfrak{su_{2}}}2 \,\, \overset{\mathfrak{g_{2}}}3  \,\, 1 \,\, \overset{\mathfrak{f_{4}}}5 \,\,  1 \,\,  ... [E_8]~. \label{quivD5}
\eea
This example illustrates a phenomenon which appears in several other flows: an $\mathfrak{e_8}$ gets broken to $\mathfrak{e_7}$, while at the same time 4 tensors are lost. Indeed $\Delta n_T=4$ receives two contributions, namely $2$ from the difference between the numbers of tensors multiplets appearing in quivers \eref{quivE6a3} and \eref{quivD5}, and $2$ from the fact that the maximal $\overset{\mathfrak{e_{8}}}{12}$ curve of $E_8$ group becomes $\overset{\mathfrak{e_{8}}}{10}$ in \eref{quivE6a3}. Amusingly, all the large contributions from the gauge group dimensions and the tensor multiplets cancel out almost completely, giving:
\bea
\Delta d  &=  99-100 = -1 ~,  \\
\quad 29 \Delta n_T + \Delta n _H -\Delta n _V &= (4\times 29) - (248-133) = 1 ~. \nn
\eea

\item  $(\CO_L, \CO'_L) = (D_7(a_2), A_7)$. The corresponding F-theory quivers are
\bea
D_7(a_2): &\qquad {\overset{\mathfrak{e_{6}}}6}  \,\, 1  \,\, \overset{\mathfrak{su_{2}}}2 \,\,\overset{\mathfrak{g_{2}}}3  \,\, 1 \,\, \overset{\mathfrak{f_{4}}}5 \,\,  1 \,\,  ... [E_8] \\
A_7: & \qquad  {\overset{\mathfrak{f_{4}}}5}  \,\, 1   \,\,  \underset{[Sp(1)]}{\overset{\mathfrak{g_{2}}}3}  \,\, 1 \,\, \overset{\mathfrak{f_{4}}}5 \,\,  1  \,\, \overset{\mathfrak{g_{2}}}3 \,\, \overset{\mathfrak{su_{2}}}2 \,\, 2 \,\, 1\,\, \overset{\mathfrak{e_{8}}}{12}  \,\, 1 \,\,  ... [E_8]
\eea
In this final case, again large group dimensions are involved, which cancel out almost completely:
\bea
\Delta d  &=  108-109 = -1 ~,  \\
\quad 29 \Delta n_T + \Delta n _H -\Delta n _V &= (1\times 29) + (8-7)- (78+3-52) = 1 ~.\nn
\eea
\ei

\subsection{The coefficient $\gamma$}
The anomaly coefficient $\gamma$ of the theory associated with the theory $T_{\CO_L, \CO_R}$ of gauge group $G$ is given by
\bea
\gamma = \frac{1}{24 \cdot 240}\left[ 30 \mathfrak{n} + 7 (\dim(G) + 1 - d_{\CO_L} - d_{\CO_R})  \right]~. \label{gammaquiv}
\eea
From \eref{dimTHG}, this can be rewritten as
\bea
\gamma = \frac{1}{24 \cdot 240} \left[ 7 (29n_T + n_H- n_V) -180 \mathfrak{n}\right]~. \label{gammaalt}
\eea
Let us discuss the application of formula \eref{gammaquiv} in various examples below.

\paragraph{The T-brane theories with $G=SU(k)$.} For quivers \eref{quivSU} consisting only of special unitary groups, formula \eref{gammaalt} simply reduces to
\bea
\gamma = \frac{1}{24 \cdot 240} \left[ 7 d_H  +23 (N-1) \right] =  \frac{1}{24 \cdot 240} \left[ \frac{7}{2} \sum_{i=1}^{N-1} f_i r_i + 30 (N-1) \right]~,
\eea
where we have used the fact that $\mathfrak{n} =n_T= N-1$ (\ie~ the number of gauge groups in the quiver) and that the dimension of the Higgs moduli space $d_H  = n_H-n_V$ is given by \eref{dHSU}. This is in agreement with (3.9) of \cite{Cremonesi:2015bld}.

\paragraph{The theories on $Q$ M5-branes on $\BC^2/\Gamma$.}  For these theories, $\mathfrak{n}=Q-1$ and $d_\CO =0$. It follows from \eref{gammaquiv} that
\bea
\gamma = \frac{1}{5760} \left[ 30(Q-1) + 7\left( \dim(G) +1 \right) \right]~,
\eea
where $\dim(G)$ is the dimension of group $G$ that is in McKay correspondence to $\Gamma$.

\paragraph{The minimal nilpotent orbit.} As before, we set $\CO_L = \CO$ and $\CO_R = 0$ for the following examples.
The coefficient $\gamma$ with $\CO$ taken to be the minimal nilpotent orbit (\ie~ that with the Bala--Carter label $A_1$ with dimension $h_G^\vee-1$) of group $G$ is given by
\bea
\gamma_{\text{min}} = \frac{1}{5760} \left[ 30 \mathfrak{n} + 7\left( \dim(G) -h^\vee_G+2 \right) \right]~,
\eea
where $h^\vee_G$ is the dual Coxeter number of group $G$ and in this case $\mathfrak{n}$ is the number of times that the gauge group $G$ appears in the quiver.

\paragraph{The subregular and the principal orbits.}  We take $\CO_R=0$ and $\CO_L=\CO$ to be the subregular and principal orbits, respectively. For $G=E_6, \, E_7, \, E_8$, these orbits are denoted by the Bala--Carter labels $G(a_1)$ and $G$ respectively, whose dimensions are given by \eref{dimsubregprinc}.  From \eref{gammaquiv}, the anomaly coefficients $\gamma$ for these theories are given by
\bea
\gamma_{\text{subreg}} &= \frac{1}{24 \cdot 240}\left[ 30 \mathfrak{n} + 7 \left (\frac{1}{2}\dim(G) + \frac{1}{2} \mathrm{rank}(G) \right)  \right]~, \\
\gamma_{\text{reg}} &= \frac{1}{24 \cdot 240}\left[ 30 \mathfrak{n} + 7 \left (\frac{1}{2}\dim(G) + \frac{1}{2} \mathrm{rank}(G) + 1 \right)  \right]~.
\eea

\subsection{The coefficient $\beta$}
The anomaly coefficient $\beta$ of the theory associated with nilpotent orbit $\CO_L=\CO, \CO_R= 0$ of group $G$ is
\bea
\beta= \frac{1}{24} \left[ -b_G (N_{\rm gauge}+1) +\frac{1}{2} N_{\rm gauge} -\frac{1}{2} n_V + B_{G} (\CO) \right]~, \label{coeffbeta}
\eea
where $N_{\rm gauge}$ is the number of gauge groups in the quiver; $n_V$ is the total dimension of the gauge groups in the quiver; $b_G$ is a rational number that depends only on the group $G$, not on the orbit and not on $N_{\rm gauge}$;  $B_G(\CO)$ depends on the orbit $\CO$ for a given group $G$ and not on $N_{\rm gauge}$.  The values of $b_G$ for various groups are as follows:
\bea
b_G = \begin{cases} 0 & \quad \text{for $G= SU(k)$} \\
3(k-3) & \quad \text{for $G=SO(2k)$} \\
\frac{41}{2} & \quad \text{for $G=E_6$} \\
\frac{113}{4} & \quad \text{for $G=E_7$} \\
\frac{125}{2} & \quad \text{for $G=E_8$}
\end{cases}
\eea

The parameter $B_G(\CO)$ has the following properties:
\bi
\item For $G=SU(k)$, $B_{SU(k)}(\CO)=0$ for all orbits $\CO$.
\item For any group $G$, $B_{G}(0)=0$ for the trivial orbit $\CO=0$.
\item For exceptional groups $G= E_6, \, E_7, \, E_8$, we have $B_G(A_1)=h^\vee_G -1$ for the minimal orbit $A_1$.
\item For the subregular and the principal orbits of $G=E_6, \, E_7, \, E_8$, we have
\bea
\begin{array}{ll}
B_{E_6}(E_6(a_1)) = 56 & \qquad B_{E_6}(E_6) = \frac{113}{2} \\
B_{E_7}(E_7(a_1)) = \frac{179}{2} &\qquad B_{E_7}(E_7) = 90 \\
B_{E_8}(E_8(a_1)) = 266 &\qquad B_{E_8}(E_8) = \frac{533}{2}~.
\end{array}
\eea
In general, we observe that
\bea
B_G(\text{principal}) -B_G(\text{subregular})  = \frac{1}{2}~.
\eea
\ei
We do not have a rule to compute $B_{G}(\CO)$ for a general orbit $\CO$.  Nevertheless, in the subsequent section, we discuss a method that allows one to compute $\beta$ for all orbits $\CO$ of $G=SO(2k)$ and, furthermore, we provide the tables that contain $\beta$ for all orbits of $E_6$, $E_7$ and $E_8$ in Section \ref{sec:athm}.

Let us discuss the application of \eref{coeffbeta} in various examples.

\paragraph{The T-brane theories of type $G=SU(k)$.} For quivers \eref{quivSU} consisting only of special unitary groups, formula \eref{coeffbeta} simply reduces to
\bea
\beta = \frac{1}{24} \left[\frac{1}{2}(N-1) -\frac{1}{2} \sum_{i=1}^\ell (r_i^2-1) \right]=  \frac{1}{24} \left[ (N-1) -\frac{1}{2} \sum_{i=1}^\ell r_i^2 \right]~.
\eea
This is in agreement with \cite[Eq.~(3.9)]{Cremonesi:2015bld}.  Recall that the $r_i$ are given in terms of the partition as explained in (\ref{eq:si}).
With this understanding, \cite[Eq.~(3.9)]{Cremonesi:2015bld} 
can be used to express $\alpha$ and $\beta$ simply in terms of the partition $\lambda$ labelling the nilpotent orbit.

\paragraph{The theories on $Q$ M5-branes on $\BC^2/\Gamma$.}  These theories correspond to the trivial orbit $\CO=0$, hence $B_{G} (\CO)=0$ for all $G$.  Using information in Table \ref{tab:QM5}, we obtain (see also (3.23) of \cite{Ohmori:2014kda})
\bea
48 \beta  = \left[ 2- |\Gamma_G| (\mathrm{rank}(G) +1) \right]Q + (\dim(G)-1)~,
\eea
where $|\Gamma_G|$ is the order of the discrete group $\Gamma_G$, which is in the McKay correspondence to $G$; $ \mathrm{rank}(G)$ is the rank of the group $G$; and $\dim(G)$ is the dimension of $G$.  Note that
\bea
&|\Gamma_{SU(k)}| = k~, \quad  |\Gamma_{D_k}| = 4k-8~, \quad |\Gamma_{E_6} |= 24, \nn \\
& |\Gamma_{E_7}| = 48, \qquad |\Gamma_{E_8}| = 120~.
\eea
Explicitly, these are
\bea
48 \beta = \begin{cases}
\left(-4 k^2+4 k+10\right) Q+[k(2k-1)-1]~, &\qquad G=SO(2k) \\
- 166 Q+77 & \qquad G=E_6 \\
- 382 Q+132 & \qquad G=E_7 \\
-1078 Q+ 247 & \qquad G= E_8~.
\end{cases}
\eea

\paragraph{The minimal nilpotent orbit.} The coefficient $\beta$ for the minimal nilpotent orbit (\ie~ that with the Bala--Carter label $A_1$) of group $G$ is given by
\bea
48 \beta  = \left[ 2- |\Gamma_G| (\mathrm{rank}(G) +1) \right](\mathfrak{n}+1) + [\dim(G)+2(h^\vee_G-1)]~,
\eea
where $h^\vee_G$ is the dual Coxeter number of group $G$.

\paragraph{The subregular and the principal orbits.}  For $G=E_6, \, E_7, \, E_8$, these orbits are denoted by the Bala--Carter labels $G(a_1)$ and $G$ respectively.  The anomaly coefficients $\beta$ for these theories can be computed using the F-theory quivers given in \cite{Heckman:2016ssk} along with \eref{coeffbeta}.  The results are as follows. 
\bea
-48\beta_{G(a_1)} =  
\begin{cases}
 {166\fn -332}~, &\qquad G=E_6 \\
 {382\fn -900}~,& \qquad G=E_7 \\
 {1078\fn -2930}~,& \qquad G=E_8
\end{cases}
\eea
and 
\bea
-48\beta_{G} =  
\begin{cases}
166\fn -449~, &\qquad G=E_6 \\
382\fn -1283~,& \qquad G=E_7 \\
1078\fn -4009~,& \qquad G=E_8
\end{cases}  
\eea
These results are also summarized in Tables \ref{tab:abE6}, \ref{tab:abE7} and \ref{tab:abE8}.  Note that for these theories, $\mathfrak{n}$ and the number of times $n_{\rm max}$ that the `maximal curves' of a given gauge group $G$ ($\overset{\mathfrak{so}_{2k}}{4}$, $\overset{\mathfrak{e_{6}}}{6}$, $\overset{\mathfrak{e_{7}}}{8}$ or $\overset{\mathfrak{e_{8}}}{12}$) appears in the F-theory quiver are related as follows:
\bea
\mathfrak{n} = n_{\rm max}+ \begin{cases} 3  & ~\text{for $E_6(a_1)$} \\  3 &~\text{for $E_7(a_1)$}\\ 4 &~\text{for $E_8(a_1)$} 
\end{cases}
\eea
and
\bea
 \mathfrak{n} = n_{\rm max}+\begin{cases} 4 &~\text{for $E_6$} \\
4 &~\text{for $E_7$} \\
5 &~\text{for $E_8$}~.
\end{cases}
\eea

\paragraph{The formula in terms of the dimension of the orbit.} 
From these examples, we make the following empirical observation:
\bea \label{coeffbeta2}
\beta=\frac{1}{48} \Big[ \left[ 2- |\Gamma_G| (\mathrm{rank}(G) +1) \right] (\mathfrak{n}+1) + \dim(G) +2d_{\CO} -1 + \Lambda_G (\CO)\Big]~,
\eea
where $\Lambda_G(\CO)$ is an integer that depends only on the orbit $\CO$ of a given group $G$.  (Note that this function is different from $B_{G}(\CO)$ in the above formulae). It has the following properties:
\bi
\item $ \Lambda_G(0)=0$ for the trivial orbit $\CO=0$.
\item $ \Lambda_G(A_1)=1$ for the minimal nilpotent orbit $\CO =A_1$.
\item $ \Lambda_G(\text{principal}) -  \Lambda_G(\text{subregular}) =  |\Gamma_G| (\mathrm{rank}(G) +1)-3$ ~.
\ei
For $\CO$ the principal orbit, the values of $\Lambda_{G}(\text{principal})$ for $G= E_6,\; E_7,\; E_8$ are
\bea
 \Lambda_{G}(\text{principal})  = \begin{cases} 
516 &\qquad G= E_6~, \\
1407 &\qquad G= E_7~, \\
4600 &\qquad G=E_8~.
\end{cases}
\eea 

\section{T-brane theories of type $G=SO(2k)$} \label{sec:SO(2n)}
\subsection{Anomaly coefficients of orthosymplectic linear quivers}
Let us consider the following quiver diagram:
{\large
\bea \label{quivTDsigrho}
\node{\Ver{}{f_1}}{p_1} -\Node{\ver{}{g_1}}{q_1}-\cdots-\node{\Ver{}{f_{i}}}{p_{i}}-\Node{\ver{}{g_{i}}}{q_{i}}-\cdots-\node{\Ver{}{f_{\ell}}}{p_{\ell}} -\Node{\ver{}{g_{\ell}}}{q_{\ell}}
\eea}where each gray node with a label $p_i$ denotes an $SO(p_i)$ group and each black node with a label $q_i$ denotes a $USp(q_i)$ group.  Here $p_i \in \BZ_{>0},~ q_i \in 2\BZ_{>0}$ for $i=1, \ldots, \ell$ and we also allow the possibility that $p_1=f_1=0$.

Let us denote by $N-1$ the number of independent tensor multiplets, or equivalently the total number of gauge groups in quiver \eref{quivTDsigrho},
\bea
N-1 = 2\ell \quad \text{or} \quad  2\ell-1~,
\eea
where the latter corresponds to the case in which $p_1=f_1=0$.

The contribution $I^{\text{vec}}$ of the vector multiplets to the anomaly polynomial $I_8$ is
\bea
I^{\text{vec}} &= -\frac{1}{24} \sum_{i=1}^\ell \Bigg[ (p_i-8) \tr F_i^4 + 3 (\tr F_i^2)^2 + 6(p_i-2) c_2(R) \tr F_i^2 + \frac{1}{2} p_i (p_i-1) c_2(R)^2 + \nn \\
& \qquad \frac{1}{2} p_1(T) \Big \{ (p_i-2) \tr F_i^2 + \frac{1}{2} p_i (p_i-1) c_2(R) \Big \} +\frac{1}{480} p_i(p_i-1) \Big \{ 7 p_1(T)^2 - 4 p_2 (T) \Big \} \Bigg] + \nn\\ 
& -\frac{1}{24} \sum_{i=1}^\ell \Bigg[ (q_i+8) \tr \hat{F}_i^4 + 3 (\tr \hat F_i^2)^2 + 6(q_i+2) c_2(R) \tr \hat F_i^2 + \frac{1}{2} q_i (q_i+1) c_2(R)^2 + \nn \\
& \qquad \frac{1}{2} p_1(T) \Big \{ (q_i+2) \tr \hat F_i^2 + \frac{1}{2} q_i (q_i+1) c_2(R) \Big \} +\frac{1}{480} q_i(q_i+1) \Big \{ 7 p_1(T)^2 - 4 p_2 (T) \Big \} \Bigg]~,
\eea
where $\tr$ denotes the trace in the fundamental representation of the corresponding group; $F_i$ and $\hat F_i$ are the field strengths of the $SO(p_i)$ and $USp(q_i)$ gauge groups, respectively.

The contribution $I^{\text{bif}}$ of the bifundamental hypermultiplets to $I_8$ is
\bea
2 I^{\text{bif}} &= \frac{1}{24} \sum_{i=1}^\ell \Bigg[ p_i \tr F_i^4 + q_i \tr \hat F_i^4 +6 \tr F_i^2 \tr \hat F_i^2 + \frac{1}{2} p_1(T) \left( p_i \tr \hat F_i^2 + q_i \tr F_i^2 \right)  \nn \\
& +\frac{p_i q_i}{240} \{ 7 p_1(T)^2 - 4 p_2 (T) \} \Bigg] 
+ \frac{1}{24} \sum_{i=1}^\ell \Bigg[ q_i \tr F_{i+1}^4 + p_{i+1} \tr \hat F_i^4 +6 \tr F_i^2 \tr \hat F_{i+1}^2  \nn \\
& + \frac{1}{2} p_1(T) \left( q_i \tr \hat F_{i+1}^2 + p_{i+1} \tr F_i^2 \right)  +\frac{q_i p_{i+1}}{240} \{ 7 p_1(T)^2 - 4 p_2 (T)  \} \Bigg]~.
\eea

The contribution $I^{\text{fund}}$ of the fundamental hypermultiplets to $I_8$ is
\bea
2 I^{\text{fund}} &= \frac{1}{24} \sum_{i=1}^\ell \left( f_i \tr F_i^4 + \frac{1}{2} f_i p_1(T) \tr F_i^2 + \frac{f_i p_i}{240} \{ 7 p_1(T)^2 - 4 p_2 (T)  \} \right)+ \nn \\
& \frac{1}{24} \sum_{i=1}^\ell \left( g_i \tr F_i^4 + \frac{1}{2} g_i p_1(T) \tr F_i^2 + \frac{g_i q_i}{240} \{ 7 p_1(T)^2 - 4 p_2 (T)  \} \right)
\eea

The contribution $I^{\text{tens}}$ of the tensor multiplets to $I_8$ is
\bea
I^{\text{tens}} &= \frac{1}{24}(N-1) \left[  c_2(R)^2 + \frac{1}{2} c_2(R) p_1(T) +\frac{1}{240}  \{ 23 p_1(T)^2 - 116 p_2 (T)  \}  \right]~.
\eea

The sum of all of the above contributions is then
\bea
24 I^{\text{1-loop}} &= 24 (I^{\text{vec}} + I^{\text{bif}} + I^{\text{fund}} + I^{\text{tens}}) \nn \\
&=  \sum_{i=1}^\ell \left[ \left(8-p_i + \frac{1}{2} \left( q_i+q_{i-1} +f_i \right)  \right) \tr F_i^4 +  \left(-8-q_i + \frac{1}{2} \left( p_i+p_{i+1} +g_i \right)  \right) \tr \hat F_i^4 \right] \nn\\
&+ \sum_{i=1}^\ell \left[ 6(2-p_i) c_2(R) + \frac{1}{2} p_1(T) \left \{ 2-p_i +\frac{1}{2} (q_i +q_{i-1}+f_i) \right \} \right] \tr F_i^2 \nn\\
&+ \sum_{i=1}^\ell \left[ 6(-2-q_i) c_2(R) + \frac{1}{2} p_1(T) \left \{-2-q_i +\frac{1}{2} (p_i +p_{i+1}+g_i) \right \} \right] \tr F_i^2 + \nn \\
&-\frac{3}{2} \sum_{I, J=1}^{N-1} C_{IJ} \frak{t}_I \frak{t}_J + \left( N-1 + n_V \right )c_2(R)^2 + \frac{1}{2}\left( N-1 + n_V \right ) c_2(R) p_1(T)  \nn \\
&+\frac{1}{240} \left[ 7 \left(  n_H  -n_V \right) +23(N-1) \right] p_1(T)^2  \nn \\
&- \frac{1}{60} \left[ \left(  n_H -n_V \right) +29(N-1)\right] p_2(T)~,
\eea
where $n_V$ is the total dimension of gauge groups
\bea
n_V = \sum_{i=1}^\ell \left[\frac{1}{2}p_i (p_i-1)+\frac{1}{2}q_i (q_i+1) \right]~,
\eea
$n_H$ is the number of hypermultiplets
\bea
n_H =  \frac{1}{2} \sum_{i=1}^\ell  \left( p_i q_i + q_i p_{i+1} +f_i p_i +g_i q_i \right)~,
\eea
and $C_{IJ}$ are elements of the Cartan matrix of the $A_{N-1}$ algebra (with 
 $I, J =1, \ldots, N-1$) and
\bea
\frak{t}_I = \begin{cases} \tr F_i^2~, &\quad  I = 2i \\ \tr \hat{F}_i^2~, &\quad  I = 2i-1 \end{cases}~.
\eea

Gauge anomaly cancellation requires that
\bea
8-p_i + \frac{1}{2} \left( q_i+q_{i-1} +f_i \right) &=0~, \\
-8-q_i + \frac{1}{2} \left( p_i+p_{i+1} +g_i \right) &= 0~.
\eea
Hence, the above expression can be simplified to
\bea
24 I^{\text{1-loop}}
&= \sum_{i=1}^\ell  \left[ 6(2-p_i) c_2(R) - 3 p_1(T)  \right] \tr F_i^2 + \sum_{i=1}^\ell \left[ 6(-2-q_i) c_2(R) +3 p_1(T)  \right] \tr \hat{F}_i^2  \nn \\
&-\frac{3}{2} \sum_{I, J=1}^{N-1} C_{IJ} \frak{t}_I \frak{t}_J + \left( N-1 + n_V \right )c_2(R)^2 + \frac{1}{2}\left( N-1 + n_V \right ) c_2(R) p_1(T)  \nn \\
&+\frac{1}{240} \left[ 7 \left(n_H -n_V  \right)+23(N-1) \right] p_1(T)^2 \nn \\
&- \frac{1}{60} \left[  \left(n_H -n_V  \right) +29(N-1)\right] p_2(T)~,
\eea
The first three terms on the right-hand side contain $\tr F_i^2$, $\tr \hat F_i^2$ and $\tr F_i^2 \hat F_i^2$.   The gauge anomaly in these terms can also be cancelled by the Green--Schwarz--West mechanism.  In order to do so, we rewrite these terms as
\bea \label{completesq}
& -\frac{3}{2} \sum_{I, J} C_{IJ} \frak{t}_I \frak{t}_J   + 6 \sum_{I} \frak{a}_I \frak{t}_I \nn \\
&= -\frac{3}{2} \sum_{I, J} C_{IJ} (\frak{t}_I -2 \frak{b}_I)( \frak{t}_J  -2 \frak{b}_J)   + 6 \sum_{I,J} C^{-1}_{IJ} \frak{a}_I \frak{a}_J
\eea
where 
\bea
\frak{a}_I &= \begin{cases} (2-p_i) c_2(R)-\frac{1}{2} p_1(T)~&\qquad I = 2i~, \\  (-2-q_i) c_2(R)+\frac{1}{2} p_1(T)~&\qquad I = 2i-1  \end{cases}~, \\
\frak{b}_I  &= \sum_{K=1}^{N-1} C^{-1}_{IK} \frak{a}_K~.
\eea
The first term of \eref{completesq} is an inner product with the bilinear form $C_{IJ}$.  This suggests that the appropriate Green--Schwarz term is
\bea
24 I^{\text{GS}} = \frac{3}{2} \sum_{I, J} C_{IJ} (\frak{t}_I -2 \frak{b}_I)( \frak{t}_J  -2 \frak{b}_J)~.
\eea
Thus, the required anomaly polynomial is
\bea
24 I^{\text{tot}} &= 24 I^{\text{1-loop}}+ 24 I^{\text{GS}} \nn \\
&= 6 \sum_{I,J} C^{-1}_{IJ} \frak{a}_I \frak{a}_J+ \left( N-1 - n_V \right )c_2(R)^2 + \frac{1}{2}\left( N-1 - n_V \right ) c_2(R) p_1(T)  \nn \\
&+\frac{1}{240} \left[ 7  \left( n_H  -n_V\right) +23(N-1) \right] p_1(T)^2 \nn \\
& - \frac{1}{60} \left[  \left( n_H  -n_V\right) +29(N-1)\right] p_2(T)~.
\eea

If the number of gauge groups is odd (\ie~$N-1$ is odd), this expression can be rewritten as
\bea
I^{\text{tot}} &= \alpha c_2(R)^2 + \beta  c_2(R) p_1(T)  + \gamma p_1(T)^2 + \delta p_2(T)~, \label{Itotorthosym}
\eea
where 

\bea
\alpha &=\frac{1}{24}\left[ 6 \sum_{I,J=1}^{N-1} C_{IJ}^{-1} R_I R_J +   N - 1 - n_V \right]   \label{anomcoeffDnalpha} \\ 
\beta &=\frac{1}{24}\left[  -3\sum_{L=1}^{N/2} R_{2L-1} +\frac{1}{2}(N-1)-\frac{1}{2} n_V \right]    \label{anomcoeffDnbeta} \\
\gamma &=\frac{1}{24} \Bigg[ \frac{3}{8} N + \frac{1}{240} \Big\{ 7  \left(n_H-n_V  \right)+23(N-1) \Big\} \Bigg]   \label{anomcoeffDngamma} \\
\delta &= -\frac{1}{24 \cdot 60} \left[ n_H -n_V + 29 (N-1) \right]~. \label{anomcoeffDndelta}
\eea
with
\bea
R_I = \begin{cases} p_i-2 &\quad \text{if $I=2i$, labelling the $SO(p_i)$ gauge group}~, \\ q_i+2 &\quad \text{if $I=2i-1$, labelling the $USp(q_i)$ gauge group}~.  \end{cases}
\eea
and $I, J= 1, \ldots, N-1$.

Note that the numbers $R_I$ for a group $G$, where $G$ is either $SO(p)$ or $USp(q)$, have a group theoretic interpretation as the ratio between the trace of $F^2$ in the adjoint representation and that in the fundamental representation.  This ratio for the group $G$ is written as $h^\vee_G/s_G$ in \cite{Ohmori:2014kda}.  Note also that it is easy to relate $p_i$ and $q_i$ for the quiver to the partition that labels the corresponding nilpotent orbit using (\ref{condrj}), so that (\ref{anomcoeffDnalpha})-(\ref{anomcoeffDndelta}) express the anomal polynomial coefficients in terms of nilpotent orbit data.

As an example, let us apply formula \eref{Itotorthosym} to compute the anomaly polynomial for the theory on $Q$ M5-branes probing $\BC^2/\hat{D}_k$ singularity.  The quiver for this theory is given by
\bea
\Node{\ver{}{2k}}{2k-8} -\node{}{2k}-\Node{}{2k-8}-\cdots -\node{}{2k} -\Node{\ver{}{2k}}{2k-8}
\eea
where there are $Q-1$ $SO(2k)$ gauge groups and $Q$ $USp(2k-8)$ gauge groups.  In this example,
\bea
N-1=2Q-1~, \qquad R_I = \begin{cases} 2k-6 &\quad \text{$I$ odd} \\ 2k-2 &\quad \text{$I$ even} \end{cases}~.
\eea
The total dimension of the gauge groups is given by
\bea
n_V &= (Q-1) k(2k-1) + Q(k-4)(2k-7) =  k [4 (k-4) Q-2 k+1]+28 Q ~.
\eea
From \eref{anomcoeffDnalpha}--\eref{anomcoeffDndelta}, we obtain the anomaly polynomial coefficients
\bea
\alpha &= \frac{1}{24} |\Gamma_{D_k}|^2 Q^3 - \frac{1}{12} Q \left[ |\Gamma_{D_k}| (k+1) -1 \right] + \frac{1}{24} [\dim(SO(2k)) -1]~,\\
\beta &= \frac{Q}{48} \left[2-  |\Gamma_{D_k}|(k+1) \right] +  \frac{1}{48} [\dim(SO(2k)) -1]~, \\
\gamma &= \frac{1}{5760} \left[ 30(Q-1) +7 \left\{ \dim(SO(2k)) +1 \right \} \right] \\
\delta &= -\frac{1}{1440} \left[ 30(Q-1) +  \dim(SO(2k)) +1 \right]~.
\eea
where $|\Gamma_{D_k}|=4k-8$ is the order of the dihedral group $\hat{D}_{k}$, which is in the McKay correspondence with the group $SO(2k)$ and the dimension of $SO(2k)$ is $\dim(SO(2k)) = k(2k-1)$.  These formulae agree with (3.23)--(3.24) of \cite{Ohmori:2014kda} after subtracting the contribution from the center-of-mass tensor multiplet, and for $Q=1$ they reproduce the anomaly coefficients of the $(D_k, D_k)$ conformal matter theory \cite{DelZotto:2014hpa} given in (3.19) of \cite{Ohmori:2014kda}.   Note that $\beta$, $\gamma$ and $\delta$ are also in agreement with the results in the preceding section.    

\subsection{The formal Type IIA construction} \label{sec:formalquiv}
In this subsection we present the Type IIA brane construction, along the lines of \cite{Hanany:1997gh, Brunner:1997gk}, for the theories corresponding to the nilpotent orbits of $SO(2k)$.  As pointed out in \cite{Heckman:2016ssk}, for some orbits there is a peculiarity in such a brane set-up, namely the presence of a non-positive number of branes in a suspended brane configuration.  Below we discuss the condition in which this peculiarity comes up and demonstrate that, despite this oddity, one can use such a `formal' brane set-up to compute anomaly coefficients of the T-brane theory in question.

Recall that nilpotent orbits of $G=SO(2k)$ are labeled by ``even partitions of $2k$," which are partitions of $2k$ subject to the constraint that every even number must appear an even number of times.  Let $\lambda$ be an even partition of $2k$ and let $\lambda^T = [\hat{\lambda}_1, \hat{\lambda}_2, \ldots, \hat{\lambda}_n]$, with $\hat{\lambda}_1 \geq \hat{\lambda}_2 \geq \cdots \geq \hat{\lambda}_n >0$, be the transpose of $\lambda$.  We define 
\bea 
\rho_i \equiv \hat{\lambda}_{i} - \hat{\lambda}_{i+1}~, \qquad i=1, \ldots, n~,
\eea
with $\rho_n \equiv \hat{\lambda}_n$.  The quiver corresponding to the partition $\lambda$ is given by\footnote{See also (6.4) of \cite{Benini:2010uu} and (2.7) of \cite{Cremonesi:2014uva}.}
\bea \label{quivTbraneSOSp}
{\large\Node{\ver{}{\rho_{1}}}{r_{1}}-\node{\Ver{}{\rho_{2}}}{r_{2}}-\Node{\ver{}{\rho_3}}{r_3}-\node{\Ver{}{\rho_{4}}}{r_{4}}-\cdots}
\eea
where each black node with the label $r_i= q_{2i-1}$ represents $\mathfrak{usp}(q_{2i-1})$ gauge algebra and each gray node with the label $r_i=p_{2i}$ represents an $\mathfrak{so}(p_{2i})$ gauge algebra. The numbers $r_j$ are fixed by
\bea \label{condrj}
r_j = \begin{cases}  -8+ \sum_{i=1}^j \hat{\lambda}_i   &\quad \text{if $j$ is odd} \\ 
\sum_{i=1}^j \hat{\lambda}_i   &\quad \text{if $j$ is even}
\end{cases}
\eea
The Type IIA brane construction of quiver \eref{quivTbraneSOSp} is \cite{Hanany:1997gh, Brunner:1997gk} similar to the one in figure \ref{fig:suk}:
\bea
\begin{tikzpicture}[baseline]
\draw[fill=black] (0,1) circle (0.3cm);
\draw[fill=black] (2,1) circle (0.3cm);
\draw[fill=black] (4,1) circle (0.3cm);
\draw[fill=black] (6,1) circle (0.3cm);
\draw[fill=black] (8,1) circle (0.3cm);
\draw [dashed,red,very thick] (0+0.3,1)--(2-0.3,1) node[black,midway, xshift =0cm, yshift=-1.5cm]{};
\draw [dashed,blue,very thick] (2+0.3,1)--(4-0.3,1) node[black,midway, xshift =0cm, yshift=-1.5cm]{};
\draw [dashed,red,very thick] (4+0.3,1)--(6-0.3,1) node[black,midway, xshift =0cm, yshift=-1.5cm]{};
\draw [dashed,blue,very thick] (6+0.3,1)--(8-0.3,1) node[black,midway, xshift =0cm, yshift=-1.5cm]{};
\node at (9,1) {$\cdots$};
\draw [solid,black] (1,0)--(1,2.5) node[black,midway, xshift =0cm, yshift=-1.5cm] {\footnotesize $\rho_{1}$};
\draw [solid,black] (3,0)--(3,2.5) node[black,midway, xshift =0cm, yshift=-1.5cm] {\footnotesize $\rho_{2}$};
\draw [solid,black] (5,0)--(5,2.5) node[black,midway, xshift =0cm, yshift=-1.5cm] {\footnotesize $\rho_{3}$};
\draw [solid,black] (7,0)--(7,2.5) node[black,midway, xshift =0cm, yshift=-1.5cm] {\footnotesize $\rho_{4}$};
\draw[solid,black,very thick] (0,1.05)--(2,1.05) node[black,midway, xshift =0.3cm, yshift=0.2cm] {\footnotesize $r_{1}$} ;  
\draw[solid,black,very thick] (2,1.05)--(4,1.05) node[black,midway, xshift =0.3cm, yshift=0.2cm] {\footnotesize $r_{2}$}; 
\draw[solid,black,very thick] (4,1.05)--(6,1.05) node[black,midway, xshift =0.3cm, yshift=0.2cm] {\footnotesize $r_{3}$};
\draw[solid,black,very thick] (6,1.05)--(8,1.05) node[black,midway, xshift =0.3cm, yshift=0.2cm] {\footnotesize $r_{4}$};
\end{tikzpicture}  
\eea
The black solid circle indicates the NS5-branes; the vertical line with the label $\rho_i$ indicates $\rho_i$ D8-branes; the red/blue dashed line indicates the $\mathrm{O6}^{+/-}$ plane and and the horizontal black solid line with the label $r_i$ indicates $r_i$ D6-branes and their image. 

Note that quiver \eref{quivTbraneSOSp} satisfies the anomaly cancellation condition which requires that $USp(N_c)$ and $SO(N_c)$ gauge theories with $N_f$ flavors must fulfil the relations $N_f =N_c+8$ and $N_f=N_c-8$ respectively \cite{Hanany:1997gh}, \ie~
\bea
r_{2j-1} &= \frac{1}{2}(r_{2j-2}+r_{2j}+ \rho_{2j-1}) +8~, \nn \\
r_{2j} &= \frac{1}{2}(r_{2j-1}+r_{2j+1}+ \rho_{2j}) -8~.
\eea

From \eref{condrj}, we see that $r_{2j-1}$ is negative or zero if and only if
\bea
\sum_{i=1}^{2j-1} \hat{\lambda}_i  \leq 8~.
\eea
Since $\hat{\lambda}_1 \geq \hat{\lambda}_2 \geq \cdots \geq \hat{\lambda}_k >0$, quiver \eref{quivTbraneSOSp} contains a gauge group with non-positive rank if and only if $\hat{\lambda}_1 \leq 8$.  In other words, we reach the following conclusion:
\begin{quote}
{\it Given a nilpotent orbit of $SO(2k)$ in terms of a D-partition of $2k$, the IIA construction has a non-positive number of branes in a suspended brane configuration if and only if the largest part of the transpose of such a partition is less than or equal to $8$.}
\end{quote}
We shall henceforth refer to such a brane configuration as the ``formal'' IIA brane construction and the corresponding quiver diagram as the ``formal'' quiver.  It should be emphasized that the quiver obtained from the IIA construction coincides with the F-theory quiver only if the ranks of all gauge groups in the quiver are non-negative\footnote{In this case, the gauge group $USp(0)$ in the IIA construction corresponds to the $(-1)$-curve in the F-theory quiver.}.  On the other other hand, the formal quiver containing gauge groups with negative ranks are different from the F-theory quiver, as all gauge groups in the latter are positive.  Note that the F-theory quiver in the latter situation contains matter in a spinor representation, which does not appear in the formal quiver obtained from the IIA construction \cite{Heckman:2016ssk}. 


Although a formal quiver does not provide a good physical description of the theory, it is a very useful tool for computing anomaly coefficients of the theory corresponding to a given nilpotent orbit.  Indeed, we have explicitly checked for all nilpotent orbits of $SO(10)$ and $SO(12)$ that the anomaly coefficients $\alpha$, $\beta$, $\gamma$ and $\delta$ computed for a formal quiver using \eref{anomcoeffDnalpha} and \eref{anomcoeffDndelta} agree with the results of \cite{Ohmori:2014kda} applied to the corresponding F-theory quiver.  Moreover, $\gamma$ and $\delta$ agree with those obtained by applying the universal formulae \eref{deltaquiv} and \eref{gammaquiv} to the F-theory quiver.  This is important because it allows us to describe the anomaly polynomial coefficients in terms of the partition labelling the nilpotent orbit via (\ref{condrj}), even when the F-theory quiver contains spinor representations.  We demonstrate this in the example below.

\paragraph{Example.} Let us consider the principal orbit $[2k-1,1]$ of $SO(2k)$.  The formal and F-theory quivers  for this theory are respectively
{\small
\bea
\Node{\ver{}{1}}{-6}-\node{}{3}-\Node{}{-4}-&\node{}{5}-\Node{}{-2}-\node{}{7}-\Node{}{0} - \node{}{9} - \Node{}{2} -  \cdots -  \node{}{2k-1} - \Node{\ver{}{1}}{2k-8}  -  \node{}{2k} - \Node{\ver{}{2k}}{2k-8}   \label{formalquiv55} \\
&  2 \quad \, \, \overset{\mathfrak{su_2}}2 \quad  \overset{\mathfrak{g_{2}}}3  \quad   1 \quad  \overset{\mathfrak{so_{9}}}4  \quad   \overset{\mathfrak{usp_2}}1  \,\,\,\,    \cdots \,\,  \overset{\mathfrak{so}_{2k-1}}4  \,\,\,\,  \underset{\left[N_f=\frac{1}{2} \right]}{\overset{\mathfrak{usp}_{2k-8}}1} \,\,\,\,   \overset{\mathfrak{so}_{2k}}4 \,\,\,\,  \overset{\mathfrak{usp}_{2k-8}}1 \,\,\,\, [SO(2k)]  \label{Ftheoryquiv55}
\eea}

In the following, we compare the anomaly coefficients of the formal quiver \eref{formalquiv55} and those of the F-theory quiver \eref{Ftheoryquiv55} in the case of $k=5$:
\bea
\Node{\ver{}{1}}{-6}-\node{}{3}-\Node{}{-4}-&\node{}{5}-\Node{}{-2}-\node{}{7}-\Node{}{0}-\node{}{9}-\Node{\ver{}{1}}{2}-\node{}{10}-\Node{\ver{}{10}}{2} \label{partform}~, \\
&2 \quad \, \, \overset{\mathfrak{su_2}}2 \quad  \overset{\mathfrak{g_{2}}}3  \quad   1 \quad \overset{\mathfrak{so}_{9}}4  \,\,\,\,  \underset{\left[N_f=\frac{1}{2} \right]}{\overset{\mathfrak{usp}_{2}}1} \,\,   \overset{\mathfrak{so}_{10}}4 \,\,\,\,  \overset{\mathfrak{usp}_{2}}1  \,\,\,\, [SO(10)]~. \label{partF}
\eea 

For $\delta$, we apply \eref{anomcoeffDndelta} to \eref{partform} and obtain
\bea
\delta = -\frac{1}{24 \cdot 60} \left[ -143 + (29 \times 11) \right]  = -\frac{11}{90}~.
\eea
On the other hand, applying \eref{deltaquiv} to \eref{partF} yields
\bea
\delta = -\frac{1}{24 \cdot 60} \left[ 29 \times 8 + 48-104 \right] =  -\frac{11}{90}~.
\eea
Thus, the two approaches give the same results.

For $\gamma$, applying \eref{anomcoeffDngamma} to \eref{partform} and applying \eref{gammaalt} to \eref{partF} yield, respectively
\bea
\gamma&=  \frac{1}{24} \left[ \left( \frac{3}{8} \times 12 \right) + \frac{1}{240} \left[ -143 +(23 \times 11) \right]  \right] = \frac{83}{1440}~, \\
\gamma&=  \frac{1}{24 \cdot 240} \left[ 7 ( 29 \times 8 + 48 -104) -180 \times 5  \right]= \frac{83}{1440}~.
\eea

For $\alpha$ and $\beta$, applying \eref{anomcoeffDnalpha} and \eref{anomcoeffDnbeta} to \eref{partform}, we obtain
\begin{align}
\alpha =  \frac{1}{24} \left[ 11 -143 +( 6 \times 1420) \right]= \frac{699}{2}~, \\
\beta = \frac{1}{24} \left[ \left( \frac{1}{2} \times 11 \right) -\frac{143}{2} -12  \right] = -\frac{13}{4}~.
\end{align}
These agree with the results for the quiver \eref{formalquiv55} computed via the usual prescription of \cite{Ohmori:2014kda}, and $\beta$ agrees with \eref{coeffbeta} upon setting $B_{SO(10)}(9,1) = 13$:
\begin{equation}
\beta = \frac{1}{24}\left[-6 \times 7 + \frac{1}{2} \times 6  - \frac{1}{2} \times 104 + 13  \right] = - \frac{13}{4}~.
\end{equation}

\subsubsection{The classification}
In this subsection, we classify all quivers that admit a formal Type IIA construction.  As discussed before, these formal quivers can be obtained by applying \eref{condrj} to the D-partitions of $2k$ whose largest part of the transpose is less than or equal to $8$.  The anomaly coefficient $\delta$ can be computed using \eref{anomcoeffDndelta}.  This can then be used to systematically construct the F-theory quiver that has the same anomaly coefficient.  Indeed, as we see below the structure of such F-theory quivers is consistent with those presented in \cite{Heckman:2016ssk}. 

There are nine cases to be considered.  In the following, we present only the part of the quiver tail that is relevant to the given partition.  Explicit examples from \cite{Heckman:2016ssk} are provided for the consistency check.
\ben
\item  $\lambda^t = [8,m,n, \ldots]$ and $m \geq n$.  The formal quiver is
\bea
\Node{\ver{}{8-m}}{0}~-~\node{\Ver{}{m-n}}{8+m}-\Node{}{m+n}-\cdots
\eea
The F-theory quiver is
\bea
[SO(8-m)] \,\, 1\,\,  \underset{[USp(m-n)]}{\overset{\mathfrak{so(8+m)}}4}  \,\, \overset{\mathfrak{usp(m+n)}}1 \,\, \ldots  
\eea
Example: $m=3, n=1; \lambda= (3,2^2,1^5)$.
\bea
[SO(5)] \,\, 1\,\,  \underset{[Sp(1)]}{\overset{\mathfrak{so(11)}}4}  \,\, \overset{\mathfrak{usp(4)}}1 \,\, \overset{\mathfrak{so(12)}}4  \,\,  \overset{\mathfrak{usp(4)}}1  \,\, ... [SO(12)]
\eea
\item  $\lambda^t = [6,m,n, \ldots]$ with $1< n \leq m \leq 6$ and $m-n$ even.  The formal quiver is
\bea
\Node{\ver{}{6-m}}{-2}~-~\node{\Ver{}{m-n}}{6+m}-\Node{}{m+n-2}-\cdots
\eea
The F-theory quiver is
\bea
[USp(m-n)] \,\,  {\overset{\mathfrak{so(6+m)}}3}   \,\, \overset{\mathfrak{usp(m+n-2)}}1 \,\, \ldots~,  
\eea
where the matter between $[USp(m-n)]$ and  ${\overset{\mathfrak{so}(6+m)}3}$ transforms under the representation $\frac{1}{2}(\mathbf{m-n}, \mathbf{s})$ of $USp(m-n) \times SO(6+m)$, where $\mathbf{s}$ denotes the spinor representation of $SO(6+m)$. \\~\\
Example: $m=3, n=3; \lambda= (3^3,1^3)$.
\bea
[Sp(1)] \,\,   {\overset{\mathfrak{so(9)}}3}  \,\, \underset{[SO(3)]}{\overset{\mathfrak{usp}(4)}1} \,\, \overset{\mathfrak{so}(12)}4  \,\,  \overset{\mathfrak{usp}(4)}1  \,\, ... [SO(12)]
\eea
\item  $\lambda^t = [6,1,1, \ldots]$.  The formal quiver is
\bea
\Node{\ver{}{5}}{-2}~-~\node{}{7}-\Node{}{0}-\node{}{9}-\cdots
\eea
The F-theory quiver is
\bea
[USp(4)] \,\,  {\overset{\mathfrak{so(7)}}3}   \,\, 1 \,\, \overset{\mathfrak{so(9)}}4 \,\, \ldots~,
\eea
where the matter between $[USp(4)]$ and  ${\overset{\mathfrak{so(7)}}3}$ transform under the representation $\frac{1}{2}(\mathbf{4},\mathbf{8})$ of $USp(4) \times SO(7)$.\\~\\
Example: $\lambda= (7,1^5)$.
\bea
[Sp(2)] \,\, \overset{\mathfrak{so}(7)}3  \,\, 1 \,\, {\overset{\mathfrak{so}(9)}4} \,\, {\overset{\mathfrak{usp}(2)}1} \,\, {\overset{\mathfrak{so}(11)}4}  \,\,  \overset{\mathfrak{usp}(4)}1  \,\, ... [SO(12)]
\eea
\item  $\lambda^t = [4,m,n, \ldots]$ with $m+n >4$, $m \geq n$ and $m-n$ even.  There are actually three possibilities, namely $(m,n)=(4,4)$, $(4,2)$ or $(3,3)$.
The formal quiver is
\bea
\Node{\ver{}{4-m}}{-4}~-~\node{\Ver{}{m-n}}{4+m}-\Node{}{m+n-4}-\cdots
\eea
The F-theory quiver is
\bea
[USp(m-n)] \, \, \,  \overset{\mathfrak{g}}3 \,\, {\overset{\mathfrak{usp}(m+n-4)}1} \, \, \ldots~,
\eea
where $\mathfrak{g}$ is $\mathfrak{so}_7$ if $m=4$ and $\mathfrak{g}_2$ if $m=3$. \\~\\
Example 1: $m=n=4; \lambda = (3^4)$.
\bea
\overset{\mathfrak{so}(7)}3  \,\, \underset{[SO(4)]}{\overset{\mathfrak{usp(4)}}1} \,\, {\overset{\mathfrak{so(12)}}4}  \,\,  \overset{\mathfrak{usp(4)}}1  \,\, ... [SO(12)]
\eea
Example 2: $\lambda^t = [4^2,2^2]; \lambda = [4^2, 2^2]$.
\bea
[SU(2)] \,\, \overset{\mathfrak{so}(7)}3  \,\, {\overset{\mathfrak{usp}(2)}1} \,\, \underset{[SU(2)]}{\overset{\mathfrak{so}(12)}4}  \,\,  \overset{\mathfrak{usp}(4)}1  \,\, ... [SO(12)]
\eea
Example 3: $\lambda^t = [4,3^2,2]; \lambda = [4^2,3,1]$.
\bea
  \overset{\mathfrak{g}_{2}}3  \,\, \overset{\mathfrak{usp}(2)}1 \,\, \underset{[SU(2)]}{\overset{\mathfrak{so}(12)}4}  \,\,  \overset{\mathfrak{usp}(4)}1  \,\, ... [SO(12)]
\eea
\item  $\lambda^t = [4,m,n, \ldots]$ with $2< m+n \leq 4$, $m \geq n$ and $m-n$ even.  There are actually two possibilities for $\lambda$ to be a D-partition, namely $(m,n)=(3,1)$, $(2,2)$.
The formal quiver is
\bea
\Node{\ver{}{4-m}}{-4}~-~\node{\Ver{}{m-n}}{4+m}-\Node{}{0}-\cdots
\eea
The F-theory quiver is
\bea
[USp(m-n)] \, \, \,  \overset{\mathfrak{g}}3 \,\,\, {1} \, \, \ldots~,
\eea
where $\mathfrak{g}$ is $\mathfrak{so}(7)$ if $m=4$ and $\mathfrak{su}(3)$ if $m=2$. \\~\\
Example 1: $\lambda^t = [4,3,1^5]; ~\lambda= [7,2^2,1]$.
\bea
[Sp(1)] \,\, \overset{\mathfrak{g}_{2}}3  \,\, 1 \,\, {\overset{\mathfrak{so}(9)}4} \,\, {\overset{\mathfrak{usp}(2)}1} \,\, {\overset{\mathfrak{so}(11)}4}  \,\,  \overset{\mathfrak{usp}(4)}1  \,\, ... [SO(12)]
\eea
Example 2: $\lambda^t = [4,2^2,1^4]; ~\lambda= [7,3,1^2]$.
\bea
\overset{\mathfrak{su}(3)}3  \,\, 1 \,\, {\overset{\mathfrak{so}(9)}4} \,\, {\overset{\mathfrak{usp}(2)}1} \,\, {\overset{\mathfrak{so}(11)}4}  \,\,  \overset{\mathfrak{usp}(4)}1  \,\, ... [SO(12)]
\eea
\item  $\lambda^t = [4,1,1, \ldots]$. The formal quiver is
\bea
\Node{\ver{}{3}}{-4}-\node{}{5}-\Node{}{-2}-\node{}{7}-\Node{}{0}-\node{}{9}-\Node{}{2}-\cdots
\eea
The F-theory quiver is
\bea
\overset{\mathfrak{su}(2)}2  \,\, \underset{[SU(2)]}{\overset{\mathfrak{so}(7)}3}  \,\, 1 \,\, {\overset{\mathfrak{so}(9)}4} \,\, {\overset{\mathfrak{usp}(2)}1}\,\, \ldots
\eea
Example: $\lambda^t = [4,1^8]; ~\lambda=[9,1^3]$.
\bea
\overset{\mathfrak{su}(2)}2  \,\, \underset{[SU(2)]}{\overset{\mathfrak{so}(7)}3}  \,\, 1 \,\, {\overset{\mathfrak{so}(9)}4} \,\, {\overset{\mathfrak{usp}(2)}1} \,\, {\overset{\mathfrak{so}(11)}4}  \,\,  \overset{\mathfrak{usp}(4)}1  \,\, ... [SO(12)]
\eea
\item  $\lambda^t = [2^5,m, \ldots]$, with $m=1, 2$.  The formal quiver is
\bea
\Node{}{-6}-\node{}{4}-\Node{}{-2}-\node{}{8}-\Node{}{2}-\node{\Ver{}{2\delta_{m,2}}}{10+m}-\Node{}{4}-\cdots
\eea
The F-theory quiver is
\bea
 \overset{\mathfrak{su}(2)}2 \,\, \overset{\mathfrak{so}(7)}3  \,\, \overset{\mathfrak{usp}(2)}1 \,\, \underset{[USp(2\delta_{m,2})]}{\overset{\mathfrak{so}(10+m)}4}  \,\,  \overset{\mathfrak{usp}(4)}1  \,\, ... [SO(12)]
 \eea

Example 1: $\lambda^t = [2^6]; ~\lambda=[6^2]$.
\bea
 \overset{\mathfrak{su}(2)}2 \,\, \overset{\mathfrak{so}(7)}3  \,\, \overset{\mathfrak{usp}(2)}1 \,\, \underset{[SU(2)]}{\overset{\mathfrak{so}(12)}4}  \,\,  \overset{\mathfrak{usp}(4)}1  \,\, ... [SO(12)]
\eea
Example 2: $\lambda^t= [2^5,1^2]; ~\lambda=[7,5]$.
\bea
\overset{\mathfrak{su}(2)}2  \,\, \overset{\mathfrak{so}(7)}3  \,\, {\overset{\mathfrak{usp}(2)}1} \,\, {\overset{\mathfrak{so}(11)}4}  \,\,  \overset{\mathfrak{usp}(4)}1  \,\, ... [SO(12)]
\eea
\item  $\lambda^t = [2^3,1 ,1, \ldots]$. The formal quiver is
\bea
\Node{}{-6}-\node{}{4}-\Node{}{-2}-\node{}{7}-\Node{}{0}-\node{}{9}-\Node{}{2}-\cdots
\eea
The F-theory quiver is
\bea
\overset{\mathfrak{su}(2)}2  \,\, \overset{\mathfrak{g_2}}3  \,\, 1 \,\, {\overset{\mathfrak{so}(9)}4} \,\, {\overset{\mathfrak{usp}(2)}1} \,\,  \ldots
\eea
Example: $\lambda^t= [2^3, 1^6]; \, \lambda=[9,3]$.
\bea
\overset{\mathfrak{su}(2)}2  \,\, \overset{\mathfrak{g_2}}3  \,\, 1 \,\, {\overset{\mathfrak{so}(9)}4} \,\, {\overset{\mathfrak{usp}(2)}1} \,\, {\overset{\mathfrak{so}(11)}4}  \,\,  \overset{\mathfrak{usp}(4)}1  \,\, ... [SO(12)]
\eea
\item  $\lambda^t = [2,1 ,1, \ldots]$.  We then have $\lambda = [2k-1,1]$ for $SO(2k)$. This in fact corresponds to the principal orbit.  The formal and F-theory quivers are given by \eref{formalquiv55} and \eref{Ftheoryquiv55}.
\een

\section{The $a$-theorem} \label{sec:athm}

\subsection{$G=SU(k)$}

The purpose of this section is to establish the $a$-theorem for flows between 6D SCFTs constructed from systems of D6-NS5-D8 branes, as discussed in \cite{Hanany:1997gh, Brunner:1997gk} and recently in \cite{Cremonesi:2015bld,Gaiotto:2014lca}.  Such theories take the form of linear quivers:
\begin{equation}
[N_{f}
=r_L]\underset{N-1}{\underbrace{\overset{\mathfrak{su}(r_1)}{2}%
,\overset{\mathfrak{su}(r_2)}{2}....\overset{\mathfrak{su}(r_{N-2})}{2}%
,\overset{\mathfrak{su}(r_{N-1})}{2}}}[N_{f}=r_R]
\end{equation}
where $N-1$ is the number of tensor multiplets and is also equal to the number of gauge groups.

For the purpose of calculating the anomaly polynomial, a $-2$ curve without a gauge algebra can be treated as an $\mathfrak{su}(1)$ gauge algebra, which means that we can take $r_i \geq 1$ for all $i$. The result $\Delta a > 0$ was explicitly established in \cite{Heckman:2015axa} for theories with up to 25 tensor multiplets, and extrapolating numerically to a larger number of tensor multiplets establishes the $a$-theorem for these theories beyond a reasonable doubt.  However, we now can prove the result analytically in full generality using formulae for $\Delta a$.

To do this, we first note that a necessary condition for a Higgs branch flow between two theories with $N-1$ tensor nodes is $r_i^{(UV)} \geq r_i^{(IR)}$ for all $i =1,...,N-1$.  We can formally decompose any given flow into a finite sequence of flows, each of which involves decreasing only a single $r_i$ by $1$.  Sometimes, the SCFT quivers in the intermediate stages of this process will violate the convexity condition $2 r_i \geq r_{i-1} + r_{i+1}$.  These correspond to ``bogus theories" in the language of \cite{Heckman:2015axa}, which feature negative numbers of hypermultiplets charged under gauge groups.  Nonetheless, an anomaly polynomial may be {\it formally} assigned to these bogus theories.  As long as $\Delta a$ decreases at each step, including steps involving bogus theories, it is guaranteed to decrease along the full flow.

The anomaly coefficients $\alpha$, $\beta$, $\gamma$, and $\delta$ are given by (\ref{eq:Atypechange}) \cite{Cremonesi:2015bld}.  We adopt the normalization of anomaly coefficients as in the preceding sections.  Any Higgs branch flow preserves Lorentz invariance, which as noted earlier implies $\Delta \gamma = \Delta \delta =0$ under any such flow \cite{Cordova:2015fha}.  It is important to emphasise that these conditions are satisfied only after accounting for free hypermultiplets, which generically show up at the endpoint of such a flow; hence they do not contradict formula \eref{diffdelta}.  These free hypermultiplets contribute only to $\gamma$ and $\delta$, so the values of $\Delta \alpha$ and $\Delta \beta$ are calculated simply from the interacting quiver theories in the UV and the IR.

From \cite{Cordova:2015fha}, we have
\begin{equation}
a = \frac{16}{7} (\alpha -\beta+\gamma) + \frac{6}{7} \delta.
\end{equation}
In order to establish $\Delta a > 0$, since $\Delta \gamma = \Delta \delta = 0$, it suffices to show $\Delta \alpha > 0, \Delta \beta < 0$ for any flow under which $r_j \rightarrow r_j - 1$ for some $j = 1, \ldots,N-1$.  For such flows, from (\ref{eq:Atypechange}) we have
\begin{equation}
24\Delta \beta = - \frac{1}{2} r_j^2 + \frac{1}{2} (r_j -1 )^2  = -r_j + \frac{1}{2}.
\end{equation}
Since $r_j \geq 1$, $\Delta \beta$ is manifestly negative, establishing the result we wanted.

$\Delta \alpha$ is slightly more complicated to compute.  Nevertheless, one can use the fact that (see (3.13) and (3.14) of \cite{Cremonesi:2015bld})
\bea
C^{-1}_{ij} &= \frac{1}{N} \begin{cases} i(N-j)~, & \qquad i \leq j \\ j(N-i) & \qquad i \geq j \end{cases}~, \\
\sum_{i,j=1}^{N-1} C^{-1}_{ij} r_i r_j &= \frac{1}{N} \left( \sum_{i=1}^{N-1}  i (N-i) r_i^2  + 2 \sum_{1 \leq i <j\leq N-1} i(N-j) r_i r_j \right) \label{idenCinv}
\eea
to find the following simple form:
\begin{equation}
24 \Delta \alpha = \alpha_{0, j} + \sum_{i=1}^{N-1} \alpha_{i,j} r_i, 
\label{eq:Aalphachange}
\end{equation}
with
\begin{align}
N \alpha_{0,j} &= j - (12 j-1)(N-j)  ,
\end{align}
and
\begin{align}
N \alpha_{i,j} &= \left\{ \begin{array}{cc} 
24 i (N-j) &\qquad j > i \\
2 [-j + (12 j-1)(N-j)]   & \qquad j = i \\
24 j (N-i) &\qquad  j < i 
\end{array} \right. .
\end{align}
Since $r_i \geq 1$ and $\alpha_{i,j} \geq 0$ for all $1 \leq i, j \leq N-1$, we have 
\bea
24 N \Delta \alpha &\geq  N\alpha_{0,j} +  \sum_{i=1}^{N-1} N\alpha_{i,j} \nn \\
&=12 j (N-1) (N-j)-N >0 \qquad \text{for all $1 \leq j \leq N-1$}~.
\eea
From this, we conclude that any RG flow has $\Delta \alpha > 0$, $\Delta \beta < 0$  , $\Delta \gamma = \Delta \delta =0$, thereby establishing the $a$-theorem for such flows.

\subsection{$G=SO(2k)$}
In this section, we repeat the analysis for quivers consisting of $SO(2k)$ gauge groups using the results of section \ref{sec:SO(2n)}.  The quivers take the form
\begin{equation}
[N_{f}
=r_L]\underset{N-1}{\underbrace{\overset{\mathfrak{usp}(r_1)}{1}%
,\overset{\mathfrak{so}(r_2)}{4}....\overset{\mathfrak{so}(r_{N-2})}{4}%
,\overset{\mathfrak{usp}(r_{N-1})}{1}}}[N_{f}=r_R]
\label{eq:Dtypequiver}
\end{equation}
where the number of gauge groups is $N-1$.
Note that here, $N-1$ must be odd, as we remarked above \eref{Itotorthosym}.  We define $r_{2i} = p_i$ and $r_{2i-1}=q_i$, so that the gauge groups alternate between $\mathfrak{so}(p_i)$ and $\mathfrak{usp}(q_i)$.  Here and subsequently, we take $I,J,K=1, \ldots N-1$ and $i,j,k = 1, \ldots, N/2$, with $p_{N/2}=0$.  Note that, subsequently, we shall consider also the case in which some $q_i$ are non-positive and some $p_i$ are less than 8, \ie~ quivers from the formal IIA construction discussed in Section \ref{sec:formalquiv}.

The anomaly polynomials for such a theory are given by \ref{eq:Dtypechange}.  Once again, $\Delta \gamma$ and $\Delta \delta$ vanish for Higgs branch flows, so $\Delta a$ depends solely on $\Delta \alpha$ and $\Delta \beta$.  

As before, it suffices for us to consider flows with just $r_J \rightarrow r_J -1$ for some $J$.  Note that although $\mathfrak{usp}(r_I)$ for odd $r_I$ does not make sense group-theoretically, we can regard a theory containing such a group as a ``bogus theory'' and still apply formulae \eref{eq:Dtypechange} to compute its anomaly coefficients; any physical flow involving $r_I \rightarrow r_I -2$ can be thought of as a sequence of two flows with $r_I \rightarrow r_I - 1$.

In such a case, we may write $\Delta \alpha$ as
\begin{equation}
24 \Delta \alpha = \alpha_{0, J} + \sum_{I=1}^{N-1} \alpha_{I, J} r_I.
\end{equation}
where upon using \eref{idenCinv}, we obtain
\begin{equation}
N\alpha_{0,J} = 6 J^2 + N(12 - 6 J- 11\delta_{J \text{ mod } 2}) ,
\end{equation}
with $\delta_{J \text{ mod } 2}$ is 1 if $J$ is even and 0 if $J$ is odd, and
\begin{align}
N \alpha_{I,J} &= \left\{ \begin{array}{cc} 
12 I (N-J) & \qquad J > I \\
-12 J^2 + (12 J-1)N  & \qquad J = I \\
12 J (N-I)  & \qquad J < I 
\end{array} \right.
\label{eq:DtypealphaIJ}
\end{align}
We consider first the case where neither the UV nor the IR theory has spinor representations.  In such a case, the smallest gauge groups we can have for the IR theory are $\mathfrak{so}(8)$ and $\mathfrak{usp}(0)$.  For $I$ odd, this implies $r_I \geq 0$ when $I \neq J$ and $r_I \geq 1$ when $I =J$.  For $I$ even, it implies $r_I \geq 8$ when $I \neq J$ and $r_I \geq 9$ when $I = J$.  From these bounds on $r_I$ as well as (\ref{eq:DtypealphaIJ}), we see clearly that $\alpha_{I,J} r_I$ is non-negative for all $I \neq J$.  It is not hard to show that $\alpha_{0,J} + \alpha_{J,J} r_J$ is constrained to be positive, so $\Delta \alpha$ is positive along all such flows.

It is also possible to prove $\Delta \alpha > 0$ for the flows involving spinor representations.  The anomaly polynomials for these theories may be calculated either via the usual techniques of \cite{Ohmori:2014kda} or by analytically continuing the formulae of (\ref{eq:Dtypechange}) to negative rank $r_i$, utilizing the formal Type IIA brane construction of section \ref{sec:formalquiv}.  This complicates matters because the positivity of $\alpha_{I,J}$ in (\ref{eq:DtypealphaIJ}) is no longer sufficient to demonstrate $\Delta \alpha > 0$ when some $r_I$ are negative.  However, the situation is improved by the fact that there are only a small number of configurations involving spinor representations.  This allows one to establish $\Delta \alpha > 0$ by brute force.

Once again, the proof proceeds by considering flows with $r_J \rightarrow r_J-1$, possibly involving bogus theories, with the understanding that some $r_I$ on the far left or far right of the quiver might be negative.  The proof above in the case without spinor representations shows that $\alpha_{I,J} r_I$ is positive whenever $r_I  \geq 0$ for $I$ odd, $r_I \geq 8$ for $I$ even.  This means that we only need to worry about the $\alpha_{I,J}$ with $r_I < 0$ ($I$ odd) or $r_I < 8$ ($I$ odd).  For a given flow, we thus consider the quantity
\begin{equation}
\alpha_- = \alpha_{0,J} + \sum_{I \in \mathcal{I}_-} \alpha_{I,J} r_I,
\end{equation}
where $\mathcal{I}_-$ is the set of indices with $r_I < 0$ ($I$ odd) or $r_I < 8$ ($I$ odd).  We have checked by brute force that this quantity is positive for all of the flows involving spinor representations.  For instance, let us consider a flow between the theories
\begin{equation}
\overset{\mathfrak{so}(10)}{3} \,\, \overset{\mathfrak{usp}(2)}{1} \,\, ...   \,\,\rightarrow \overset{\mathfrak{so}(9)}{3} \,\, \overset{\mathfrak{usp}(2)}{1}  \,\, ... \,\, 
\end{equation}
The anomaly polynomials of these theories may be computed via an analytic continuation of (\ref{eq:Dtypechange}) with $ r_1 = -2 $, $r_2 = 10 \rightarrow 9$.  This flow has $\alpha_- = \alpha_{0,2} - 2 \alpha_{1,2} + 10 \alpha_{2,2} =  195 N-408$, where $N-1$ is the number of tensor multiplets in the original quiver of (\ref{eq:Dtypequiver}) (i.e. before the blowdown).  $\alpha_-$ is postive whenever $N-1 \geq 2$.  Any quiver with spinor representations necessarily has $N-1 \geq 2$, since for $N-1=1$ we have no $\mathfrak{so}(r_I)$ gauge algebras in the quiver at all.  This shows that $\alpha_- > 0$ for such a flow, establishing the result.

As one flows further down the RG hierarchy, the number of tensor multiplets in the quiver $N$ required to demonstrate $\alpha_- > 0$ increases, but so does the number of tensor multiplets required to realize the flow.  For instance, the final flow in the hierarchy is
\begin{equation}
\overset{\mathfrak{su}(2)}{2} \,\,\overset{\mathfrak{g}_2}{3} \,\,    {1} \,\,\overset{\mathfrak{so}(9)}{4}  \,\, ...  \,\,\rightarrow 2 \,\, \overset{\mathfrak{su}(2)}{2} \,\,\overset{\mathfrak{g}_2}{3} \,\,    {1} \,\,\overset{\mathfrak{so}(9)}{4}   \,\, ... \,\, .
\end{equation}
The anomaly polynomial of the UV theory is given by (\ref{eq:Dtypechange}) with $(r_1,r_2,r_3,r_4,r_5,r_6) = (-6,4,-2,7,0,9)$, while the IR theory has $(r_1,r_2,r_3,r_4,r_5,r_6) = (-6,3,-4,5,-2,7)$; see \eref{formalquiv55}.  Proving $\alpha_- > 0$ for each step $r_J \rightarrow r_J -1$ flow requires us to assume $N -1\geq 6$.  This constraint is necessarily satisfied, as the presence of $r_6$ requires $N-1 \geq 6$.

$\Delta \beta$ takes a much simpler form:
\bea
\Delta \beta = \begin{cases}
-3 - \frac{1}{2} r_J  &\qquad \text{$J$ odd} \\
\frac{1}{2}-\frac{1}{2} r_J & \qquad \text{$J$ even}~.
\end{cases}
\eea
For $J$ odd, $\Delta \beta \leq 0$ provided $r_J \geq -6$.  For $J$ even, $\Delta \beta \leq 0$ provided $r_J \geq 1$.  In theories of the form (\ref{eq:Dtypequiver}), we always have $r_J \geq 1 $ for $J$ odd, $r_J \geq 9$ for $J$ even for any $r_J \rightarrow r_J-1$ flow, so $\Delta \beta$ is always negative.  In theories with $-3$ curves, we can always use the formal quiver to compute the anomaly polynomials.  However, even in the case of such a formal quiver, we always have $r_J \geq -5$ for $J$ odd and $r_J \geq 3$ for $J$ even, so $\Delta \beta < 0$ for any such flow.

\subsection{$G=E_k$}
We have also verified $\Delta \alpha > 0$, $\Delta \beta < 0$ for all of the flows in the $E_6$, $E_7$, and $E_8$ nilpotent hierarchies, for all deformations of both the left and right sides of the quiver, using the formulae shown in Appendix \ref{Eformulae}.  As in the type D case, flows further down the RG hierarchy require more tensor multiplets to establish $\Delta \alpha > 0$, but the deformations of the quiver reach further into the interior of the quiver, so the theories involve more nodes.  In Appendix \ref{Eformulae}, we have written down explicit formulae for $\alpha$ and $\beta$ for each of the corresponding theories assuming no breaking on the right.  Here, $\mathfrak{n}$ is the number of $-2$ curves in the F-theory quiver after blowing down all $-1$ curves.  As noted after \eref{dimTHG}, $\mathfrak{n}$ is constant along the flow from one orbit to another.

\section{Conclusions}\label{sec:conclusions}

The recent work \cite{Heckman:2016ssk} showed how nilpotent orbits can be used to characterize T-brane 6D SCFTs, their RG flows, and their global symmetries.  In this work, we have seen that these nilpotent orbits are also related to the anomalies and Higgs moduli spaces of the corresponding SCFTs.

It is worth noting that the class of T-brane theories we have considered here is not the only set of 6D SCFTs with connections to group theory.  As conjectured in \cite{DelZotto:2014hpa} and verified in \cite{Heckman:2015bfa}, deformations of the worldvolume theory of M5-branes simultaneously probing a $\mathbb{C}^2/\Gamma_G$ orbifold singularity and an $E_8$ wall are in one-to-one correspondence with $\mathrm{Hom}(\Gamma_G, E_8)$.  6D SCFTs with $(2,0)$ supersymmetry are in one-to-one correspondence with semisimple Lie algebras (i.e. they admit an ADE classification) \cite{Witten:1997kz, Cordova:2015vwa}, and their anomalies can be expressed in terms of the dimension, rank, and dual Coxeter number of the associated Lie algebra \cite{Ohmori:2014kda}.  Given the connection between these 6D SCFTs and elliptically fibered Calabi-Yau threefolds that are used to produce them via F-theory, we are led to a surprising correspondence between group theory and Calabi-Yau geometry, which would be interesting to study from a purely mathematical perspective.  Given the success of the above examples, it is also tempting to conjecture that the full set of 6D SCFTs and the RG flows between them should be related in some precise way to structures in group theory.

Our analysis in this paper was limited to SCFTs with quivers that were suitably long so that the nilpotent deformation on the left of the quiver did not overlap with the nilpotent deformations on the right of the quiver.  It would be interesting to try to understand how this story extends to short quivers.  There is reason to think that this task might succeed: although \cite{Heckman:2013pva} noted that some SCFTs with short quivers seem to be outliers, \cite{Morrison:2016nrt} later showed that these ``outliers" can in fact be viewed as limiting cases of 6D SCFTs with long quivers.  A slightly modified approach to the classification of 6D SCFTs might shed light on this issue and allow one to generalize our results to short quivers. 

The results of this paper add even more evidence to the already strong case that the $a$-theorem is true for RG flows between 6D SCFTs.  However, a rigorous proof is still lacking, and would be nice to have.

\acknowledgments
We would like to express our sincere thanks to Clay Cordova, Amihay Hanany, Jonathan Heckman, Hiroyuki Shimizu, and Alberto Zaffaroni for a number of useful discussions and for sharing their insights into the subject.  We also thank the Simons Center for Geometry and Physics Summer Workshop 2016 for their hospitality during this work. NM is indebted to the CERN visitor programme from October to December 2016, during which this work was finalized. NM and AT are supported in part by the INFN. NM is also supported in part by the ERC Starting Grant 637844-HBQFTNCER. TR is supported by NSF grant PHY-1067976 and by the NSF GRF under DGE-1144152. AT is also supported in part by the ERC under Grant Agreement n.~307286 (XD-STRING), and by the MIUR-FIRB grant RBFR10QS5J ``String Theory and Fundamental Interactions''.

\appendix
\section{Explicit formulae for anomaly coefficients}\label{formulae}

In this appendix, we collect the formulae for the coefficients $\alpha$, $\beta$, $\gamma$, and $\delta$ for each of the cases $G=SU(k)$, $SO(2k)$, and $E_k$.

\subsection{$G = SU(k)$}

The theories with $G=SU(k)$ take the form
\begin{equation}
[N_{f}
=r_L]\underset{N-1}{\underbrace{\overset{\mathfrak{su}(r_1)}{2}%
,\overset{\mathfrak{su}(r_2)}{2}....\overset{\mathfrak{su}(r_{N-2})}{2}%
,\overset{\mathfrak{su}(r_{N-1})}{2}}}[N_{f}=r_R]
\end{equation}
where $N-1$ is the number of tensor multiplets and is also equal to the number of gauge groups.  $\alpha$ and $\beta$ for these theories are given by
\bea
24 \alpha &= 12 \sum_{i,j=1}^{N-1} C_{ij}^{-1} r_i r_j + 2 (N-1) - \sum_{i=1}^{N-1} r_i^2\nn \\
24 \beta &= N-1-\frac{1}{2} \sum_{i=1}^{N-1} r_i^2 \nn \\
24 \gamma &= \frac{1}{240} \left( \frac{7}{2} \sum_{i-1}^{N-1} r_i f_i + 30 (N-1) \right) \nn \\
24 \delta &= -\frac{1}{120} \left( \sum_{i=1}^{N-1} r_i f_i + 60 (N-1) \right)~. 
\label{eq:Atypechange}
\eea
Here, $C_{ij}$ is the $(N-1) \times (N-1)$ Cartan matrix for $A_{N-1}$.

\subsection{$G = SO(2k)$}

The theories with $G=SO(2k)$ take the form
\begin{equation}
[N_{f}
=r_L]\underset{N-1}{\underbrace{\overset{\mathfrak{usp}(r_1)}{1}%
,\overset{\mathfrak{so}(r_2)}{4}....\overset{\mathfrak{so}(r_{N-2})}{4}%
,\overset{\mathfrak{usp}(r_{N-1})}{1}}}[N_{f}=r_R]
\label{eq:Dtypequiverapp}
\end{equation}
where the number of gauge groups is $N-1$.  Let us define $r_{2i} = p_i$ and $r_{2i-1}=q_i$, so that the gauge groups alternate between $\mathfrak{so}(p_i)$ and $\mathfrak{usp}(q_i)$.  We take $I,J,K=1, \ldots N-1$ and $i,j,k = 1, \ldots, N/2$, with $p_{N/2}=0$.  Note that the formal IIA construction discussed in Section \ref{sec:formalquiv} allows some $q_i$ to be negative and some $p_i$ to be less than 8.  $\alpha$ and $\beta$ are given for these theories by
\begin{align}
24\alpha &= 6 \sum_{I,J=1}^{N-1} C_{IJ}^{-1} r_I r_J + 12 \sum_i q_i +  7 N - 1 - n_V \nn \\ 
24\beta &= -3\sum_i (2+q_i) +\frac{1}{2}(N-1)-\frac{1}{2} n_V  \nn \\
24\gamma &=   \frac{1}{240} \left[ 7  \left(   -n_V + \frac{1}{2} \sum_{i}  ( p_i q_i + q_{i+1} p_{i} +f_i p_i +g_i q_i )  \right)+ 23 (N-1) \right]  +\frac{3}{8} N \nn \\
24\delta &= -\frac{1}{60} \left[  -n_V + \frac{1}{2} \sum_{i}  \left( p_i q_i + q_{i+1} p_{i} +f_i p_i +g_i q_i  \right) + 29 (N-1) \right]~.
\label{eq:Dtypechange}
\end{align}
Here,
\bea
 f_i = 2 p_i -16 - q_i - q_{i+1} \nn \\
 g_i = 2 q_i +16 - p_i - p_{i+1} 
\eea
\bea
n_V = \sum_{i} \left[\frac{1}{2}p_i (p_i-1)+\frac{1}{2}q_i (q_i+1) \right]~,
\eea
and $C_{IJ}$ is the Cartan matrix of rank $N-1$.

\subsection{$G = E_k$}\label{Eformulae}

$\delta$ and $\gamma$ are given by (\ref{deltaquiv}),  (\ref{deltaalt}) and (\ref{gammaquiv}), (\ref{gammaalt}) for the theories with $G= E_k$:
\bea
\begin{array}{lll}
24 \delta &= -\frac{1}{60}\left( 29 n_T + n_H - n_V\right) &=  - \frac{1}{60} \left[ 30\mathfrak{n} + \dim(G) +1- d_{\CO_L}- d_{\CO_R} \right] \\
24 \gamma &= \frac{1}{240} \left[ 7 (29n_T + n_H- n_V) -180 \mathfrak{n}\right]  &=   \frac{1}{240}\left[ 30 \mathfrak{n} + 7 (\dim(G) + 1 -  d_{\CO_L}- d_{\CO_R})  \right] 
\end{array}
\eea
where $n_T$, $n_H$, and $n_V$ are the number of tensor multiplets, hypermultiplets, and vector multiplets in the quiver description, $d_{\CO_L}$ and $d_{\CO_R}$ are the quaternionic dimension of the orbits on the left and right side of the quiver, and $\mathfrak{n}$ is the number of $-2$ curves in the F-theory quiver after blowing down all $-1$ curves. 

The formulae for $\alpha$ and $\beta$ for the theories with $\CO_L = \CO$, $\CO_R = 0$ are given in the following table.  Here, $\mathfrak{n}$ is the number of $-2$ curves in the F-theory quiver after blowing down all $-1$ curves.  In particular, for the undeformed theories (see \eg~ Examples 1 and 2 in Section \ref{sec:coeffdelta}), $\mathfrak{n}$ is simply the number of $\mathfrak{e}_6$, $\mathfrak{e}_7$, or $\mathfrak{e}_8$ gauge algebras, respectively.
\bgroup
\def\arraystretch{1.5}
\begin{longtable}{|c|c|c|}\hline
B-C Label & $\alpha$ & $\beta$ \\ \hline
$0$ & $24 \mathfrak{n}^3+72 \mathfrak{n}^2+\frac{697 \mathfrak{n}}{12}+\frac{319}{24}$&$\frac{1}{48} (-166 \mathfrak{n}-89)$ \\ \hline

$A_1$ & $\frac{\mathfrak{n} (\mathfrak{n} (288 \mathfrak{n} (\mathfrak{n}+4)+1417)+658)+99}{12 (\mathfrak{n}+1)}$&$\frac{1}{24} (-83 \mathfrak{n}-33)$ \\ \hline

$2 A_1$ & $24 \mathfrak{n}^3+72 \mathfrak{n}^2+\frac{409 \mathfrak{n}}{12}-\frac{2}{\mathfrak{n}+1}+\frac{167}{24}$&$\frac{1}{48} (-166 \mathfrak{n}-49)$ \\ \hline

$3 A_1$ & $\frac{\mathfrak{n} (\mathfrak{n} (288 \mathfrak{n} (\mathfrak{n}+4)+1129)+350)+31}{12 (\mathfrak{n}+1)}$&$\frac{1}{24} (-83 \mathfrak{n}-17)$ \\ \hline

$A_2$ & $\frac{\mathfrak{n} (3 \mathfrak{n}+2) (32 \mathfrak{n} (3 \mathfrak{n}+10)+115)+13}{12 (\mathfrak{n}+1)}$&$\frac{1}{24} (-83 \mathfrak{n}-11)$ \\ \hline

$A_2 + A_1$ & $\frac{\mathfrak{n} (2 \mathfrak{n} (288 \mathfrak{n} (\mathfrak{n}+4)+841)+267)+13}{24 (\mathfrak{n}+1)}$&$\frac{1}{48} (-166 \mathfrak{n}-11)$ \\ \hline

$A_2 + 2 A_1$ & $\frac{\mathfrak{n} (\mathfrak{n} (288 \mathfrak{n} (\mathfrak{n}+4)+697)+49)}{12 (\mathfrak{n}+1)}$&$-\frac{83 \mathfrak{n}}{24}$ \\ \hline

$2 A_2$ & $\frac{\mathfrak{n} (2 \mathfrak{n} (288 \mathfrak{n} (\mathfrak{n}+4)+409)-103)+39}{24 (\mathfrak{n}+1)}$&$\frac{5}{16}-\frac{83 \mathfrak{n}}{24}$ \\ \hline

$2 A_2 + A_1$ & $\frac{\mathfrak{n} (\mathfrak{n} (288 \mathfrak{n} (\mathfrak{n}+4)+265)-100)+13}{12 (\mathfrak{n}+1)}$&$\frac{1}{24} (13-83 \mathfrak{n})$ \\ \hline

$A_3$ & $\frac{\mathfrak{n} (2 \mathfrak{n} (288 \mathfrak{n} (\mathfrak{n}+4)+121)-253)+33}{24 (\mathfrak{n}+1)}$&$\frac{1}{48} (33-166 \mathfrak{n})$ \\ \hline

$A_3 + A_1$ & $\frac{\mathfrak{n} (\mathfrak{n} (288 \mathfrak{n} (\mathfrak{n}+4)-23)-140)+21}{12 (\mathfrak{n}+1)}$&$\frac{7}{8}-\frac{83 \mathfrak{n}}{24}$ \\ \hline

$D_4(a_1)$ & $\frac{\mathfrak{n} (\mathfrak{n} (288 \mathfrak{n} (\mathfrak{n}+4)-167)-142)+25}{12 (\mathfrak{n}+1)}$&$\frac{1}{24} (25-83 \mathfrak{n})$ \\ \hline

$A_4$ & $\frac{\mathfrak{n} (2 \mathfrak{n} (288 \mathfrak{n} (\mathfrak{n}+4)-1319)+579)+145}{24 (\mathfrak{n}+1)}$&$\frac{1}{48} (97-166 \mathfrak{n})$ \\ \hline

$A_4 + A_1$ & $\frac{\mathfrak{n} (\mathfrak{n} (288 \mathfrak{n} (\mathfrak{n}+4)-1463)+371)+52}{12 (\mathfrak{n}+1)}$&$\frac{1}{24} (52-83 \mathfrak{n})$ \\ \hline

$D_4$ & $\frac{\mathfrak{n} (2 \mathfrak{n} (288 \mathfrak{n} (\mathfrak{n}+4)-2471)+2515)-223}{24 (\mathfrak{n}+1)}$&$\frac{1}{48} (137-166 \mathfrak{n})$ \\ \hline

$D_5(a_1)$ & $\frac{\mathfrak{n} (\mathfrak{n} (288 \mathfrak{n} (\mathfrak{n}+4)-2759)+1599)-178}{12 (\mathfrak{n}+1)}$&$\frac{1}{24} (74-83 \mathfrak{n})$ \\ \hline

$A_5$ & $\frac{\mathfrak{n} (\mathfrak{n} (288 \mathfrak{n} (\mathfrak{n}+4)-3479)+2611)-396}{12 (\mathfrak{n}+1)}$&$\frac{7}{2}-\frac{83 \mathfrak{n}}{24}$ \\ \hline

$E_6(a_3)$ & $\frac{\mathfrak{n} (\mathfrak{n} (288 \mathfrak{n} (\mathfrak{n}+4)-3623)+2800)-489}{12 (\mathfrak{n}+1)}$&$\frac{1}{24} (87-83 \mathfrak{n})$ \\ \hline

$D_5$ & $\frac{\mathfrak{n} (2 \mathfrak{n} (288 \mathfrak{n} (\mathfrak{n}+4)-7079)+19535)-7779}{24 (\mathfrak{n}+1)}$&$\frac{87}{16}-\frac{83 \mathfrak{n}}{24}$ \\ \hline

$E_6(a_1)$ & $\frac{\mathfrak{n} (\mathfrak{n} (288 \mathfrak{n} (\mathfrak{n}+4)-10535)+19919)-11018}{12 (\mathfrak{n}+1)}$&$-\frac{83}{24} (\mathfrak{n}-2)$ \\ \hline
\caption{$\alpha$ and $\beta$ for the $E_6$ nilpotent hierarchy.}
\label{tab:abE6}
\end{longtable}
\egroup

\bgroup
\def\arraystretch{1.5}%
\begin{longtable}{|c|c|c|}\hline
B-C Label & $\alpha$ & $\beta$ \\ \hline

$0$ & $96 \mathfrak{n}^3+288 \mathfrak{n}^2+\frac{3073 \mathfrak{n}}{12}+\frac{835}{12}$&$\frac{1}{24} (-191 \mathfrak{n}-125)$ \\ \hline

$A_1$ & $\frac{\mathfrak{n} (2 \mathfrak{n} (1152 \mathfrak{n} (\mathfrak{n}+4)+6241)+6927)+1345}{24 (\mathfrak{n}+1)}$&$\frac{1}{48} (-382 \mathfrak{n}-215)$ \\ \hline

$2 A_1$ & $96 \mathfrak{n}^3+288 \mathfrak{n}^2+\frac{2497 \mathfrak{n}}{12}-\frac{2}{\mathfrak{n}+1}+\frac{283}{6}$&$\frac{1}{24} (-191 \mathfrak{n}-94)$ \\ \hline

$(3 A_1)'$ & $\frac{\mathfrak{n} (2 \mathfrak{n} (1152 \mathfrak{n} (\mathfrak{n}+4)+5665)+5393)+867}{24 (\mathfrak{n}+1)}$&$\frac{1}{48} (-382 \mathfrak{n}-165)$ \\ \hline

$(3 A_1)'' $ & $\frac{\mathfrak{n} (2 \mathfrak{n} (1152 \mathfrak{n} (\mathfrak{n}+4)+5665)+5413)+887}{24 (\mathfrak{n}+1)}$&$\frac{1}{48} (-382 \mathfrak{n}-169)$ \\ \hline

$A_2$ & $96 \mathfrak{n}^3+288 \mathfrak{n}^2+\frac{1921 \mathfrak{n}}{12}-\frac{8}{\mathfrak{n}+1}+\frac{295}{8}$&$\frac{1}{48} (-382 \mathfrak{n}-147)$ \\ \hline

$4 A_1$ & $\frac{(3 \mathfrak{n}+1) (\mathfrak{n} (128 \mathfrak{n} (3 \mathfrak{n}+11)+1323)+347)}{12 (\mathfrak{n}+1)}$&$\frac{1}{24} (-191 \mathfrak{n}-73)$ \\ \hline

$A_2 + A_1$ & $\frac{\mathfrak{n} (2 \mathfrak{n} (1152 \mathfrak{n} (\mathfrak{n}+4)+5089)+4109)+543}{24 (\mathfrak{n}+1)}$&$\frac{1}{48} (-382 \mathfrak{n}-129)$ \\ \hline

$A_2 +2 A_1$ & $96 \mathfrak{n}^3+288 \mathfrak{n}^2+\frac{1345 \mathfrak{n}}{12}-\frac{18}{\mathfrak{n}+1}+\frac{283}{8}$&$\frac{1}{48} (-382 \mathfrak{n}-111)$ \\ \hline

$A_2 +3 A_1$ & $\frac{\mathfrak{n} (\mathfrak{n} (1152 \mathfrak{n} (\mathfrak{n}+4)+4513)+1507)+156}{12 (\mathfrak{n}+1)}$&$-\frac{191 \mathfrak{n}}{24}-2$ \\ \hline

$2 A_2 $ & $96 \mathfrak{n}^3+288 \mathfrak{n}^2+\frac{769 \mathfrak{n}}{12}-\frac{32}{\mathfrak{n}+1}+\frac{85}{2}$&$-\frac{191 \mathfrak{n}}{24}-\frac{7}{4}$ \\ \hline

$2 A_2 + A_1$ & $\frac{\mathfrak{n} (2 \mathfrak{n} (1152 \mathfrak{n} (\mathfrak{n}+4)+3937)+2105)+171}{24 (\mathfrak{n}+1)}$&$\frac{1}{48} (-382 \mathfrak{n}-69)$ \\ \hline

$A_3$ & $\frac{\mathfrak{n} (\mathfrak{n} (1152 \mathfrak{n} (\mathfrak{n}+4)+3649)+859)+66}{12 (\mathfrak{n}+1)}$&$-\frac{191 \mathfrak{n}}{24}-\frac{5}{4}$ \\ \hline

$(A_3 + A_1)'$ & $\frac{\mathfrak{n} (2 \mathfrak{n} (1152 \mathfrak{n} (\mathfrak{n}+4)+3361)+1337)+75}{24 (\mathfrak{n}+1)}$&$\frac{1}{48} (-382 \mathfrak{n}-45)$ \\ \hline

$(A_3 + A_1)'' $ & $\frac{\mathfrak{n} (2 \mathfrak{n} (1152 \mathfrak{n} (\mathfrak{n}+4)+3361)+1381)+119}{24 (\mathfrak{n}+1)}$&$\frac{1}{48} (-382 \mathfrak{n}-49)$ \\ \hline

$A_3 + 2 A_1$ & $\frac{\mathfrak{n} (\mathfrak{n} (1152 \mathfrak{n} (\mathfrak{n}+4)+3073)+512)+31}{12 (\mathfrak{n}+1)}$&$\frac{1}{24} (-191 \mathfrak{n}-17)$ \\ \hline

$D_4(a_1)$ & $\frac{\mathfrak{n} (2 \mathfrak{n} (1152 \mathfrak{n} (\mathfrak{n}+4)+3073)+1001)+39}{24 (\mathfrak{n}+1)}$&$\frac{1}{48} (-382 \mathfrak{n}-33)$ \\ \hline

$D_4(a_1)+ A_1$ & $\frac{\mathfrak{n} (\mathfrak{n} (1152 \mathfrak{n} (\mathfrak{n}+4)+2785)+356)+13}{12 (\mathfrak{n}+1)}$&$\frac{1}{24} (-191 \mathfrak{n}-11)$ \\ \hline

$A_3 + A_2$ & $\frac{\mathfrak{n} (2 \mathfrak{n} (1152 \mathfrak{n} (\mathfrak{n}+4)+2497)+447)+13}{24 (\mathfrak{n}+1)}$&$\frac{1}{48} (-382 \mathfrak{n}-11)$ \\ \hline

$A_3 + A_2 + A_1$ & $\frac{\mathfrak{n} (\mathfrak{n} (1152 \mathfrak{n} (\mathfrak{n}+4)+2209)+103)}{12 (\mathfrak{n}+1)}$&$-\frac{191 \mathfrak{n}}{24}$ \\ \hline

$A_4$ & $\frac{\mathfrak{n} (\mathfrak{n} (1152 \mathfrak{n} (\mathfrak{n}+4)+769)-253)+34}{12 (\mathfrak{n}+1)}$&$\frac{1}{24} (22-191 \mathfrak{n})$ \\ \hline

$A_4 + A_1$ & $\frac{\mathfrak{n} (2 \mathfrak{n} (1152 \mathfrak{n} (\mathfrak{n}+4)+481)-603)+55}{24 (\mathfrak{n}+1)}$&$\frac{1}{48} (55-382 \mathfrak{n})$ \\ \hline

$A_4 + A_2$ & $\frac{\mathfrak{n} (2 \mathfrak{n} (1152 \mathfrak{n} (\mathfrak{n}+4)-383)-685)+81}{24 (\mathfrak{n}+1)}$&$\frac{1}{48} (81-382 \mathfrak{n})$ \\ \hline

$D_4$ & $\frac{\mathfrak{n} (\mathfrak{n} (1152 \mathfrak{n} (\mathfrak{n}+4)-1535)-164)+123}{12 (\mathfrak{n}+1)}$&$\frac{1}{24} (51-191 \mathfrak{n})$ \\ \hline

$D_4 + A_1$ & $\frac{\mathfrak{n} (2 \mathfrak{n} (1152 \mathfrak{n} (\mathfrak{n}+4)-1823)-233)+233}{24 (\mathfrak{n}+1)}$&$\frac{1}{48} (113-382 \mathfrak{n})$ \\ \hline

$D_5(a_1)$ & $\frac{\mathfrak{n} (2 \mathfrak{n} (1152 \mathfrak{n} (\mathfrak{n}+4)-2111)-115)+219}{24 (\mathfrak{n}+1)}$&$\frac{1}{48} (123-382 \mathfrak{n})$ \\ \hline

$D_5(a_1) + A_1$ & $\frac{\mathfrak{n} (\mathfrak{n} (1152 \mathfrak{n} (\mathfrak{n}+4)-2399)+13)+102}{12 (\mathfrak{n}+1)}$&$\frac{1}{24} (66-191 \mathfrak{n})$ \\ \hline

$A_5'' $ & $\frac{\mathfrak{n} (2 \mathfrak{n} (1152 \mathfrak{n} (\mathfrak{n}+4)-3551)+1149)+463}{24 (\mathfrak{n}+1)}$&$\frac{1}{48} (151-382 \mathfrak{n})$ \\ \hline

$A_5' $ & $\frac{\mathfrak{n} (6 \mathfrak{n}-1) (64 \mathfrak{n} (6 \mathfrak{n}+25)-917)+231}{24 (\mathfrak{n}+1)}$&$\frac{1}{48} (159-382 \mathfrak{n})$ \\ \hline

$A_5 + A_1$ & $\frac{\mathfrak{n} (\mathfrak{n} (1152 \mathfrak{n} (\mathfrak{n}+4)-3839)+670)+189}{12 (\mathfrak{n}+1)}$&$\frac{1}{24} (81-191 \mathfrak{n})$ \\ \hline

$E_6(a_3)$ & $\frac{\mathfrak{n} (2 \mathfrak{n} (1152 \mathfrak{n} (\mathfrak{n}+4)-3839)+1153)+191}{24 (\mathfrak{n}+1)}$&$\frac{1}{48} (167-382 \mathfrak{n})$ \\ \hline

$D_6(a_2)$ & $\frac{\mathfrak{n} (\mathfrak{n} (1152 \mathfrak{n} (\mathfrak{n}+4)-4415)+918)+125}{12 (\mathfrak{n}+1)}$&$\frac{1}{24} (89-191 \mathfrak{n})$ \\ \hline

$E_7 (a_5)$ & $\frac{\mathfrak{n} (\mathfrak{n} (1152 \mathfrak{n} (\mathfrak{n}+4)-4703)+1060)+93}{12 (\mathfrak{n}+1)}$&$\frac{1}{24} (93-191 \mathfrak{n})$ \\ \hline

$D_6(a_1)$ & $\frac{\mathfrak{n} (2 \mathfrak{n} (1152 \mathfrak{n} (\mathfrak{n}+4)-9599)+10223)-1299}{24 (\mathfrak{n}+1)}$&$\frac{1}{48} (285-382 \mathfrak{n})$ \\ \hline

$D_5$ & $\frac{\mathfrak{n} (\mathfrak{n} (1152 \mathfrak{n} (\mathfrak{n}+4)-10751)+6440)-953}{12 (\mathfrak{n}+1)}$&$\frac{1}{24} (151-191 \mathfrak{n})$ \\ \hline

$D_5 + A_1$ & $\frac{\mathfrak{n} (2 \mathfrak{n} (1152 \mathfrak{n} (\mathfrak{n}+4)-11039)+13523)-2139}{24 (\mathfrak{n}+1)}$&$\frac{1}{48} (309-382 \mathfrak{n})$ \\ \hline

$A_6$ & $\frac{\mathfrak{n} (2 \mathfrak{n} (1152 \mathfrak{n} (\mathfrak{n}+4)-11327)+14331)-2231}{24 (\mathfrak{n}+1)}$&$\frac{1}{48} (313-382 \mathfrak{n})$ \\ \hline

$E_7(a_4)$ & $\frac{\mathfrak{n} (\mathfrak{n} (1152 \mathfrak{n} (\mathfrak{n}+4)-11615)+7499)-1244}{12 (\mathfrak{n}+1)}$&$\frac{20}{3}-\frac{191 \mathfrak{n}}{24}$ \\ \hline

$E_6(a_1)$ & $\frac{\mathfrak{n} (2 \mathfrak{n} (1152 \mathfrak{n} (\mathfrak{n}+4)-17663)+33277)-9157}{24 (\mathfrak{n}+1)}$&$\frac{1}{48} (419-382 \mathfrak{n})$ \\ \hline

$D_6$ & $\frac{\mathfrak{n} (\mathfrak{n} (1152 \mathfrak{n} (\mathfrak{n}+4)-25151)+32109)-11884}{12 (\mathfrak{n}+1)}$&$\frac{65}{6}-\frac{191 \mathfrak{n}}{24}$ \\ \hline

$E_7(a_3)$ & $\frac{\mathfrak{n} (\mathfrak{n} (1152 \mathfrak{n} (\mathfrak{n}+4)-25439)+32730)-12301}{12 (\mathfrak{n}+1)}$&$\frac{1}{24} (263-191 \mathfrak{n})$ \\ \hline

$E_6$ & $\frac{\mathfrak{n} (\mathfrak{n} (1152 \mathfrak{n} (\mathfrak{n}+4)-38399)+67878)-36283}{12 (\mathfrak{n}+1)}$&$\frac{1}{24} (341-191 \mathfrak{n})$ \\ \hline

$E_7(a_2)$ & $\frac{\mathfrak{n} (2 \mathfrak{n} (1152 \mathfrak{n} (\mathfrak{n}+4)-39263)+141251)-76683}{24 (\mathfrak{n}+1)}$&$\frac{1}{48} (693-382 \mathfrak{n})$ \\ \hline

$E_7(a_1)$ & $96 \mathfrak{n}^3+288 \mathfrak{n}^2-\frac{63455 \mathfrak{n}}{12}-\frac{53361}{2 (\mathfrak{n}+1)}+17596$&$\frac{75}{4}-\frac{191 \mathfrak{n}}{24}$ \\ \hline

$E_7$ & $\frac{\mathfrak{n} (2 \mathfrak{n} (1152 \mathfrak{n} (\mathfrak{n}+4)-108383)+801889)-884845}{24 (\mathfrak{n}+1)}$&$\frac{1}{48} (1283-382 \mathfrak{n})$ \\ \hline

\caption{$\alpha$ and $\beta$ for the $E_7$ nilpotent hierarchy.}
\label{tab:abE7}
\end{longtable}
\egroup

\bgroup
\def\arraystretch{1.5}%
\begin{longtable}{|c|c|c|}\hline
B-C Label & $\alpha$ & $\beta$ \\ \hline
$0$ & $600 \mathfrak{n}^3+1800 \mathfrak{n}^2+\frac{20521 \mathfrak{n}}{12}+\frac{4163}{8}$&$\frac{1}{48} (-1078 \mathfrak{n}-831)$ \\ \hline

$A_1$ & $\frac{\mathfrak{n} (\mathfrak{n} (7200 \mathfrak{n} (\mathfrak{n}+4)+41401)+25541)+5734}{12 (\mathfrak{n}+1)}$&$\frac{1}{24} (-539 \mathfrak{n}-386)$ \\ \hline

$2 A_1$ & $\frac{\mathfrak{n} (2 \mathfrak{n} (7200 \mathfrak{n} (\mathfrak{n}+4)+40681)+48765)+10555}{24 (\mathfrak{n}+1)}$&$\frac{1}{48} (-1078 \mathfrak{n}-725)$ \\ \hline

$3 A_1$ & $\frac{\mathfrak{n} (\mathfrak{n} (7200 \mathfrak{n} (\mathfrak{n}+4)+39961)+23280)+4865}{12 (\mathfrak{n}+1)}$&$-\frac{49}{24} (11 \mathfrak{n}+7)$ \\ \hline

$4 A_1$ & $600 \mathfrak{n}^3+1800 \mathfrak{n}^2+\frac{17641 \mathfrak{n}}{12}-\frac{8}{\mathfrak{n}+1}+\frac{3055}{8}$&$-\frac{7}{48} (154 \mathfrak{n}+93)$ \\ \hline

$A_2$ & $\frac{\mathfrak{n} (\mathfrak{n} (7200 \mathfrak{n} (\mathfrak{n}+4)+39241)+22233)+4496}{12 (\mathfrak{n}+1)}$&$-\frac{539 \mathfrak{n}}{24}-\frac{41}{3}$ \\ \hline

$A_2 + A_1$ & $\frac{\mathfrak{n} (2 \mathfrak{n} (7200 \mathfrak{n} (\mathfrak{n}+4)+38521)+42425)+8283}{24 (\mathfrak{n}+1)}$&$\frac{1}{48} (-1078 \mathfrak{n}-621)$ \\ \hline

$A_2 + 2 A_1$ & $\frac{\mathfrak{n} (\mathfrak{n} (7200 \mathfrak{n} (\mathfrak{n}+4)+37801)+20237)+3820}{12 (\mathfrak{n}+1)}$&$-\frac{539 \mathfrak{n}}{24}-\frac{37}{3}$ \\ \hline

$A_2 + 3 A_1$ & $\frac{\mathfrak{n} (2 \mathfrak{n} (7200 \mathfrak{n} (\mathfrak{n}+4)+37081)+38593)+7043}{24 (\mathfrak{n}+1)}$&$\frac{1}{48} (-1078 \mathfrak{n}-565)$ \\ \hline

$2 A_2$ & $\frac{\mathfrak{n} (2 \mathfrak{n} (7200 \mathfrak{n} (\mathfrak{n}+4)+36361)+36781)+6491}{24 (\mathfrak{n}+1)}$&$\frac{1}{48} (-1078 \mathfrak{n}-541)$ \\ \hline

$2 A_2 + A_1$ & $\frac{\mathfrak{n} (\mathfrak{n} (7200 \mathfrak{n} (\mathfrak{n}+4)+35641)+17520)+2993}{12 (\mathfrak{n}+1)}$&$-\frac{7}{24} (77 \mathfrak{n}+37)$ \\ \hline

$A_3$ & $\frac{\mathfrak{n} (2 \mathfrak{n} (7200 \mathfrak{n} (\mathfrak{n}+4)+34921)+33409)+5567}{24 (\mathfrak{n}+1)}$&$\frac{1}{48} (-1078 \mathfrak{n}-505)$ \\ \hline

$2 A_2 + 2 A_1$ & $\frac{\mathfrak{n} (2 \mathfrak{n} (7200 \mathfrak{n} (\mathfrak{n}+4)+34921)+33347)+5505}{24 (\mathfrak{n}+1)}$&$-\frac{11}{48} (98 \mathfrak{n}+45)$ \\ \hline

$A_3 + A_1$ & $\frac{\mathfrak{n} (\mathfrak{n} (7200 \mathfrak{n} (\mathfrak{n}+4)+34201)+15872)+2545}{12 (\mathfrak{n}+1)}$&$\frac{1}{24} (-539 \mathfrak{n}-239)$ \\ \hline

$A_2 + 2 A_1$ & $\frac{\mathfrak{n} (2 \mathfrak{n} (7200 \mathfrak{n} (\mathfrak{n}+4)+33481)+30147)+4657}{24 (\mathfrak{n}+1)}$&$-\frac{7}{48} (154 \mathfrak{n}+65)$ \\ \hline

$D_4(a_1)$ & $\frac{\mathfrak{n} (2 \mathfrak{n} (7200 \mathfrak{n} (\mathfrak{n}+4)+33481)+30169)+4679}{24 (\mathfrak{n}+1)}$&$\frac{1}{48} (-1078 \mathfrak{n}-457)$ \\ \hline

$D_4(a_1) + A_1$ & $\frac{\mathfrak{n} (2 \mathfrak{n} (7200 \mathfrak{n} (\mathfrak{n}+4)+32761)+28619)+4269}{24 (\mathfrak{n}+1)}$&$\frac{1}{48} (-1078 \mathfrak{n}-435)$ \\ \hline

$A_3 + A_2$ & $\frac{\mathfrak{n} (\mathfrak{n} (7200 \mathfrak{n} (\mathfrak{n}+4)+32041)+13569)+1952}{12 (\mathfrak{n}+1)}$&$-\frac{539 \mathfrak{n}}{24}-\frac{26}{3}$ \\ \hline

$A_3 + A_2 + A_1$ & $\frac{\mathfrak{n} (2 \mathfrak{n} (7200 \mathfrak{n} (\mathfrak{n}+4)+31321)+25705)+3563}{24 (\mathfrak{n}+1)}$&$\frac{1}{48} (-1078 \mathfrak{n}-397)$ \\ \hline

$D_4(a_1) + A_2$ & $\frac{\mathfrak{n} (2 \mathfrak{n} (7200 \mathfrak{n} (\mathfrak{n}+4)+30601)+24319)+3245}{24 (\mathfrak{n}+1)}$&$\frac{1}{48} (-1078 \mathfrak{n}-379)$ \\ \hline

$A_4$ & $\frac{\mathfrak{n} (2 \mathfrak{n} (7200 \mathfrak{n} (\mathfrak{n}+4)+27721)+19313)+2271}{24 (\mathfrak{n}+1)}$&$\frac{1}{48} (-1078 \mathfrak{n}-321)$ \\ \hline

$2 A_3$ & $\frac{\mathfrak{n} (2 \mathfrak{n} (7200 \mathfrak{n} (\mathfrak{n}+4)+27721)+19247)+2205}{24 (\mathfrak{n}+1)}$&$-\frac{7}{48} (154 \mathfrak{n}+45)$ \\ \hline

$A_4 + A_1$ & $\frac{\mathfrak{n} (\mathfrak{n} (7200 \mathfrak{n} (\mathfrak{n}+4)+27001)+9060)+1013}{12 (\mathfrak{n}+1)}$&$\frac{1}{24} (-539 \mathfrak{n}-151)$ \\ \hline

$A_4 + 2 A_1$ & $\frac{\mathfrak{n} (2 \mathfrak{n} (7200 \mathfrak{n} (\mathfrak{n}+4)+26281)+16973)+1803}{24 (\mathfrak{n}+1)}$&$-\frac{539 \mathfrak{n}}{24}-\frac{95}{16}$ \\ \hline

$A_4 + A_2$ & $\frac{\mathfrak{n} (\mathfrak{n} (7200 \mathfrak{n} (\mathfrak{n}+4)+24841)+7421)+724}{12 (\mathfrak{n}+1)}$&$-\frac{539 \mathfrak{n}}{24}-\frac{16}{3}$ \\ \hline

$A_4 + A_2 + A_1$ & $\frac{\mathfrak{n} (2 \mathfrak{n} (7200 \mathfrak{n} (\mathfrak{n}+4)+24121)+13813)+1271}{24 (\mathfrak{n}+1)}$&$\frac{1}{48} (-1078 \mathfrak{n}-241)$ \\ \hline

$D_4$ & $\frac{\mathfrak{n} (2 \mathfrak{n} (7200 \mathfrak{n} (\mathfrak{n}+4)+21961)+11367)+1237}{24 (\mathfrak{n}+1)}$&$\frac{1}{48} (-1078 \mathfrak{n}-227)$ \\ \hline

$D_4 + A_1$ & $\frac{\mathfrak{n} (\mathfrak{n} (7200 \mathfrak{n} (\mathfrak{n}+4)+21241)+5197)+510}{12 (\mathfrak{n}+1)}$&$-\frac{539 \mathfrak{n}}{24}-\frac{17}{4}$ \\ \hline

$A_4 + A_3$ & $\frac{\mathfrak{n} (2 \mathfrak{n} (7200 \mathfrak{n} (\mathfrak{n}+4)+20521)+9307)+665}{24 (\mathfrak{n}+1)}$&$-\frac{7}{48} (154 \mathfrak{n}+25)$ \\ \hline

$D_5(a_1)$ & $\frac{\mathfrak{n} (\mathfrak{n} (7200 \mathfrak{n} (\mathfrak{n}+4)+20521)+4744)+423}{12 (\mathfrak{n}+1)}$&$\frac{1}{24} (-539 \mathfrak{n}-93)$ \\ \hline

$D_5(a_1) + A_1$ & $\frac{\mathfrak{n} (2 \mathfrak{n} (7200 \mathfrak{n} (\mathfrak{n}+4)+19801)+8629)+695}{24 (\mathfrak{n}+1)}$&$\frac{1}{48} (-1078 \mathfrak{n}-169)$ \\ \hline

$D_4 + A_2$ & $\frac{\mathfrak{n} (\mathfrak{n} (7200 \mathfrak{n} (\mathfrak{n}+4)+19081)+3908)+283}{12 (\mathfrak{n}+1)}$&$-\frac{77}{24} (7 \mathfrak{n}+1)$ \\ \hline

$D_5(a_1) + A_2$ & $\frac{\mathfrak{n} (2 \mathfrak{n} (7200 \mathfrak{n} (\mathfrak{n}+4)+17641)+6331)+377}{24 (\mathfrak{n}+1)}$&$\frac{1}{48} (-1078 \mathfrak{n}-127)$ \\ \hline

$A_5$ & $\frac{\mathfrak{n} (\mathfrak{n} (7200 \mathfrak{n} (\mathfrak{n}+4)+16921)+2828)+157}{12 (\mathfrak{n}+1)}$&$\frac{1}{24} (-539 \mathfrak{n}-59)$ \\ \hline

$E_6(a_3)$ & $\frac{\mathfrak{n} (\mathfrak{n} (7200 \mathfrak{n} (\mathfrak{n}+4)+16201)+2516)+139}{12 (\mathfrak{n}+1)}$&$\frac{1}{24} (-539 \mathfrak{n}-53)$ \\ \hline

$A_5 + A_1$ & $\frac{\mathfrak{n} (2 \mathfrak{n} (7200 \mathfrak{n} (\mathfrak{n}+4)+16201)+4987)+233}{24 (\mathfrak{n}+1)}$&$\frac{1}{48} (-1078 \mathfrak{n}-103)$ \\ \hline

$E_6(a_3) + A_1$ & $\frac{\mathfrak{n} (2 \mathfrak{n} (7200 \mathfrak{n} (\mathfrak{n}+4)+15481)+4387)+197}{24 (\mathfrak{n}+1)}$&$-\frac{7}{48} (154 \mathfrak{n}+13)$ \\ \hline

$D_6(a_2)$ & $\frac{\mathfrak{n} (2 \mathfrak{n} (7200 \mathfrak{n} (\mathfrak{n}+4)+14761)+3787)+137}{24 (\mathfrak{n}+1)}$&$\frac{1}{48} (-1078 \mathfrak{n}-79)$ \\ \hline

$E7(a5)$ & $\frac{\mathfrak{n} (2 \mathfrak{n} (7200 \mathfrak{n} (\mathfrak{n}+4)+14041)+3235)+101}{24 (\mathfrak{n}+1)}$&$\frac{1}{48} (-1078 \mathfrak{n}-67)$ \\ \hline

$E_8(a_7)$ & $\frac{\mathfrak{n} (2 \mathfrak{n} (7200 \mathfrak{n} (\mathfrak{n}+4)+13321)+2707)+65}{24 (\mathfrak{n}+1)}$&$-\frac{11}{48} (98 \mathfrak{n}+5)$ \\ \hline

$A_6$ & $\frac{\mathfrak{n} (\mathfrak{n} (7200 \mathfrak{n} (\mathfrak{n}+4)+1801)-919)+64}{12 (\mathfrak{n}+1)}$&$\frac{1}{24} (52-539 \mathfrak{n})$ \\ \hline

$A_6 + A_1$ & $\frac{\mathfrak{n} (2 \mathfrak{n} (7200 \mathfrak{n} (\mathfrak{n}+4)+1081)-1935)+115}{24 (\mathfrak{n}+1)}$&$\frac{1}{48} (115-1078 \mathfrak{n})$ \\ \hline

$D_5$ & $\frac{\mathfrak{n} (2 \mathfrak{n} (7200 \mathfrak{n} (\mathfrak{n}+4)-1079)-1481)+677}{24 (\mathfrak{n}+1)}$&$\frac{1}{48} (125-1078 \mathfrak{n})$ \\ \hline

$D_5 + A_1$ & $\frac{\mathfrak{n} (\mathfrak{n} (7200 \mathfrak{n} (\mathfrak{n}+4)-1799)-775)+298}{12 (\mathfrak{n}+1)}$&$-\frac{7}{24} (77 \mathfrak{n}-10)$ \\ \hline

$D_6(a_1)$ & $\frac{\mathfrak{n} (\mathfrak{n} (7200 \mathfrak{n} (\mathfrak{n}+4)-2519)-787)+268}{12 (\mathfrak{n}+1)}$&$\frac{1}{24} (76-539 \mathfrak{n})$ \\ \hline

$E_7(a_4)$ & $\frac{\mathfrak{n} (2 \mathfrak{n} (7200 \mathfrak{n} (\mathfrak{n}+4)-3239)-1551)+499}{24 (\mathfrak{n}+1)}$&$\frac{1}{48} (163-1078 \mathfrak{n})$ \\ \hline

$D_5 + A_2$ & $\frac{\mathfrak{n} (\mathfrak{n} (7200 \mathfrak{n} (\mathfrak{n}+4)-3959)-752)+231}{12 (\mathfrak{n}+1)}$&$\frac{1}{24} (87-539 \mathfrak{n})$ \\ \hline

$D_7(a_2)$ & $600 \mathfrak{n}^3-\frac{1799 \mathfrak{n}}{12}-\frac{1}{2 \mathfrak{n}}+\frac{889}{24}$&$\frac{1}{48} (325-1078 \mathfrak{n})$ \\ \hline

$E_6(a_1)$ & $\frac{\mathfrak{n} (\mathfrak{n} (7200 \mathfrak{n} (\mathfrak{n}+4)-18359)+2640)+263}{12 (\mathfrak{n}+1)}$&$\frac{1}{24} (167-539 \mathfrak{n})$ \\ \hline

$A_7$ & $\frac{\mathfrak{n} (2 \mathfrak{n} (7200 \mathfrak{n} (\mathfrak{n}+4)-18359)+4999)+245}{24 (\mathfrak{n}+1)}$&$-\frac{11}{48} (98 \mathfrak{n}-31)$ \\ \hline

$E_6(a_1) + A_1$ & $\frac{\mathfrak{n} (2 \mathfrak{n} (7200 \mathfrak{n} (\mathfrak{n}+4)-19079)+5759)+417}{24 (\mathfrak{n}+1)}$&$\frac{1}{48} (345-1078 \mathfrak{n})$ \\ \hline

$E_8(b_6)$ & $\frac{\mathfrak{n} (2 \mathfrak{n} (7200 \mathfrak{n} (\mathfrak{n}+4)-21239)+7405)+155}{24 (\mathfrak{n}+1)}$&$-\frac{7}{48} (154 \mathfrak{n}-53)$ \\ \hline

$D_6$ & $\frac{\mathfrak{n} (2 \mathfrak{n} (7200 \mathfrak{n} (\mathfrak{n}+4)-37079)+25913)-1929}{24 (\mathfrak{n}+1)}$&$\frac{1}{48} (519-1078 \mathfrak{n})$ \\ \hline

$E_7(a_3)$ & $\frac{\mathfrak{n} (2 \mathfrak{n} (7200 \mathfrak{n} (\mathfrak{n}+4)-37799)+26895)-2159}{24 (\mathfrak{n}+1)}$&$\frac{1}{48} (529-1078 \mathfrak{n})$ \\ \hline

$D_7(a_1)$ & $\frac{\mathfrak{n} (\mathfrak{n} (7200 \mathfrak{n} (\mathfrak{n}+4)-38519)+13950)-1195}{12 (\mathfrak{n}+1)}$&$\frac{1}{24} (269-539 \mathfrak{n})$ \\ \hline

$E_8(a_6)$ & $\frac{\mathfrak{n} (2 \mathfrak{n} (7200 \mathfrak{n} (\mathfrak{n}+4)-44279)+36907)-4135}{24 (\mathfrak{n}+1)}$&$\frac{1}{48} (593-1078 \mathfrak{n})$ \\ \hline

$E_6$ & $\frac{\mathfrak{n} (2 \mathfrak{n} (7200 \mathfrak{n} (\mathfrak{n}+4)-70199)+92015)-16419}{24 (\mathfrak{n}+1)}$&$\frac{1}{48} (789-1078 \mathfrak{n})$ \\ \hline

$E_6 + A_1$ & $\frac{\mathfrak{n} (\mathfrak{n} (7200 \mathfrak{n} (\mathfrak{n}+4)-70919)+46823)-8552}{12 (\mathfrak{n}+1)}$&$\frac{50}{3}-\frac{539 \mathfrak{n}}{24}$ \\ \hline

$E7(a_2)$ & $\frac{\mathfrak{n} (\mathfrak{n} (7200 \mathfrak{n} (\mathfrak{n}+4)-72359)+48511)-9216}{12 (\mathfrak{n}+1)}$&$17-\frac{539 \mathfrak{n}}{24}$ \\ \hline

$E8(b5)$ & $\frac{\mathfrak{n} (\mathfrak{n} (7200 \mathfrak{n} (\mathfrak{n}+4)-73079)+49373)-9548}{12 (\mathfrak{n}+1)}$&$\frac{1}{24} (412-539 \mathfrak{n})$ \\ \hline

$D_7$ & $\frac{\mathfrak{n} (2 \mathfrak{n} (7200 \mathfrak{n} (\mathfrak{n}+4)-88919)+141829)-34621}{24 (\mathfrak{n}+1)}$&$\frac{1}{48} (947-1078 \mathfrak{n})$ \\ \hline

$E_8(a_5)$ & $\frac{\mathfrak{n} (2 \mathfrak{n} (7200 \mathfrak{n} (\mathfrak{n}+4)-90359)+146185)-36169}{24 (\mathfrak{n}+1)}$&$\frac{7}{48} (137-154 \mathfrak{n})$ \\ \hline

$E_7(a_1)$ & $\frac{\mathfrak{n} (2 \mathfrak{n} (7200 \mathfrak{n} (\mathfrak{n}+4)-124199)+263709)-85025}{24 (\mathfrak{n}+1)}$&$-\frac{7}{48} (154 \mathfrak{n}-169)$ \\ \hline

$E_8(b_4)$ & $\frac{\mathfrak{n} (\mathfrak{n} (7200 \mathfrak{n} (\mathfrak{n}+4)-124919)+133196)-43229}{12 (\mathfrak{n}+1)}$&$-\frac{7}{24} (77 \mathfrak{n}-85)$ \\ \hline

$E_8(a_4)$ & $\frac{\mathfrak{n} (2 \mathfrak{n} (7200 \mathfrak{n} (\mathfrak{n}+4)-159479)+413703)-164939}{24 (\mathfrak{n}+1)}$&$\frac{1}{48} (1405-1078 \mathfrak{n})$ \\ \hline

$E_7$ & $\frac{\mathfrak{n} (\mathfrak{n} (7200 \mathfrak{n} (\mathfrak{n}+4)-245159)+443854)-244593}{12 (\mathfrak{n}+1)}$&$\frac{1}{24} (927-539 \mathfrak{n})$ \\ \hline

$E_8(a_3)$ & $\frac{\mathfrak{n} (\mathfrak{n} (7200 \mathfrak{n} (\mathfrak{n}+4)-245879)+446059)-246462}{12 (\mathfrak{n}+1)}$&$\frac{1}{24} (930-539 \mathfrak{n})$ \\ \hline

$E_8(a_2)$ & $\frac{\mathfrak{n} (2 \mathfrak{n} (7200 \mathfrak{n} (\mathfrak{n}+4)-332279)+1502837)-1034205}{24 (\mathfrak{n}+1)}$&$\frac{753}{16}-\frac{539 \mathfrak{n}}{24}$ \\ \hline

$E_8(a_1)$ & $\frac{\mathfrak{n} (\mathfrak{n} (7200 \mathfrak{n} (\mathfrak{n}+4)-505079)+1534298)-1404623}{12 (\mathfrak{n}+1)}$&$\frac{1}{24} (1465-539 \mathfrak{n})$ \\ \hline
\caption{$\alpha$ and $\beta$ for the $E_8$ nilpotent hierarchy.}
\label{tab:abE8}
\end{longtable}
\egroup

\bibliographystyle{ytphys}
\bibliography{6Dtabflow,at,6Dtabflow2}

\end{document}